\documentclass[useAMS,usenatbib]{mn2e}
\usepackage{graphics}
\usepackage{times}

\newcommand{\ms}{m\,s$^{-1}$} 
\newcommand{\kms}{km\,s$^{-1}$}   
\newcommand{\kmq}{km$^2$\,s$^{-2}$}   
\newcommand{\ion}[2]{#1\,{\sc #2}}
\newcommand{\te}{T$_{\rm eff}$}
\newcommand{\lgg}{$\log g$}
\newcommand{\vp}{$V_p$}
\newcommand{\vs}{$v_{\rm e}\sin i$}
\newcommand{\bs}{$\langle B \rangle$}
 
\newcommand{\equ}{$\gamma$~Equ}
\newcommand{\cir}{$\alpha$~Cir}
\newcommand{\nd}{\ion{Nd}{iii}}
\newcommand{\ndt}{\ion{Nd}{ii}}
\newcommand{\pr}{\ion{Pr}{iii}}
\newcommand{\tb}{\ion{Tb}{iii}}
\newcommand{\ha}{H$\alpha$}

\newcommand{\va}{$\langle V \rangle$}
\newcommand{\vb}{$\langle V^2 \rangle$}

\newcommand{\beq}{\begin{equation}}
\newcommand{\eeq}{\end{equation}}
\newcommand{\dfrac}[2]{\frac{\displaystyle #1}{\displaystyle #2}}

\newcommand{\fifps}[3]{\centering\resizebox{#1}{!}{\rotatebox{#2}{\includegraphics{#3}}}}

\newcommand{\figps}[1]{\resizebox{\hsize}{!}{\rotatebox{0}{\includegraphics{#1}}}}

\begin{document}

\title[Line profile variations in rapidly oscillating Ap stars]
{Line profile variations in rapidly oscillating Ap stars: \\ 
resolution of the enigma\thanks{Based on observations collected at the European Southern Observatory
(Paranal, La Silla), at the Canada-France-Hawaii Telescope and on data
retrieved from the ESO Science Archive.}}
\author
[O. Kochukhov et al.]
{O. Kochukhov$^1$\thanks{E-mail: oleg@astro.uu.se}, 
 T. Ryabchikova$^{2,3}$,
 W. W. Weiss$^3$,
 J. D. Landstreet$^4$ and
 D. Lyashko$^5$ \\
$^1$Department of Astronomy and Space Physics, Uppsala University, Box 515, 751 20 Uppsala, Sweden \\
$^2$Institute of Astronomy, Russian Academy of Sciences, Pyatnitskaya 48, 119017 Moscow, Russia \\
$^3$Department of Astronomy, University of Vienna, T\"urkenschanzstra{\ss}e 17, 1180 Vienna, Austria \\
$^4$Department of Physics and Astronomy, University of Western Ontario, London, Ontario N6A 3K7, Canada \\
$^5$Tavrian National University, Yaltinskaya 4, 95007 Simferopol, Crimea, Ukraine
} 

\date{Accepted 2006 December 28. Received 2006 December 23; in original form
2006 November 06}

\maketitle
\pubyear{2006}

\begin{abstract}  
We have carried out the first survey of the pulsational line profile variability in rapidly
oscillating Ap (roAp) stars. We analysed high signal-to-noise time-series observations of ten sharp-lined roAp 
stars obtained with the high-resolution spectrographs attached to the VLT and CFHT
telescopes. We investigated in detail the variations of \ion{Pr}{iii}, \ion{Nd}{ii}, \ion{Nd}{iii} and 
\ion{Tb}{iii} lines and discovered a prominent change of the profile variability pattern 
with height in the atmospheres of all studied roAp stars. We show that, in every investigated star,
profile variability of at least one rare-earth ion is characterized by unusual blue-to-red
moving features, which we previously discovered in the time-resolved spectra of the
roAp star $\gamma$~Equ. This behaviour is common in rapidly rotating non-radial pulsators but
is inexplicable in the framework of the standard oblique pulsator model of slowly rotating roAp
stars. Using analysis of the line profile moments and spectrum synthesis calculations,
we demonstrate that unusual oscillations in spectral lines of roAp stars arise
from the pulsational modulation of line widths.  This variation occurs
approximately in quadrature with the radial velocity changes, and its amplitude
rapidly increases with height in stellar atmosphere. We propose that the line width
modulation is a consequence of the periodic expansion and compression of turbulent
layers in the upper atmospheres of roAp stars. Thus, the line profile changes observed in
slowly rotating magnetic pulsators should be interpreted as a superposition of two
types of variability: the usual time-dependent velocity field due to an oblique
low-order pulsation mode and an additional line width modulation, synchronized with the
changes of stellar radius. Our explanation of the line profile variations of roAp stars
solves the long-standing observational puzzle and 
opens new possibilities for constraining geometric and physical properties of the
stellar magnetoacoustic pulsations. 
\end{abstract}


\section{Introduction}
\label{intro}

About 10\,\% to 20\,\% of upper main sequence stars are characterized by remarkably rich line
spectra, often containing numerous unidentified features. Compared to the solar case,
overabundances of up to a few dex are often inferred for some iron-peak and rare-earth
elements, whereas some other chemical elements are found to be underabundant 
\citep{RNW04}. Some of these \textit{chemically peculiar} (Ap) stars also exhibit
organized magnetic fields with a typical strength of a few kG. Chemical peculiarities are
believed to result from the influence of the magnetic field on the diffusing ions, possibly in
combination with the influence of a weak, magnetically-directed wind 
\citep[e.g.,][]{Babel92}.

More than 30 cool Ap stars exhibit high-overtone, low-degree, non-radial $p$-mode pulsations with
periods in the range of 6--21 minutes \citep{KM00}. These \textit{rapidly
oscillating Ap} (roAp) stars are key objects for asteroseismology, which presently is the most
powerful tool for testing theories of stellar structure and evolution. The observed pulsation
amplitudes of roAp stars are modulated according to the visible magnetic  field structure,
pointing to a defining role played by magnetic fields in exciting the oscillations and shaping the
main pulsation properties. Observation of the coincidence of the magnetic field and pulsation 
amplitude extrema gave rise to the {\it oblique pulsator model} \citep[OPM,][]{K82}. According to
this phenomenological framework, the main characteristics of non-radial pulsations in roAp stars can
be attributed to an oblique axisymmetric dipole ($\ell=1$, $m=0$) mode, aligned with the axis of
a nearly axisymmetric quasi-dipolar magnetic field. 

Several theoretical investigations \citep[e.g.,][]{BD02,SG04,S05} studied the effects of the
distortion of oblique pulsation mode geometry by the global magnetic field and stellar rotation.
\citet{BD02} predicted that the rotational distortion of pulsation eigenmodes is represented by a
superposition of spherical harmonic functions, containing substantial non-axisymmetric
components, and that there is no alignment of the pulsation axis and the dipolar magnetic field. On the
other hand, \citet{SG04} and \citet{S05} disregarded effects of rotation, focusing on the detailed
treatment of the interaction between $p$-modes and a strong magnetic field. These authors find an
axisymmetric pulsation structure and alignment between the magnetic field and the pulsation modes, and predict
a strong confinement of pulsation amplitude to the magnetic field axis.

Recently major progress in the observational study of roAp stars was achieved by employing high
time resolution spectroscopy in addition to the classical high-speed photometric measurements.
Time-series spectroscopy of magnetic pulsators revealed a surprising diversity in
the pulsational behaviour of different lines \citep[e.g.,][and references
therein]{KH98,KEM06}. Detailed analyses of the bright roAp star \equ\ \citep{SMR99,KR01a}
demonstrated that spectroscopic pulsational variability is dominated by the lines of rare-earth
ions, especially those of Pr and Nd, which are strong and numerous in the roAp spectra. On the
other hand, light and iron-peak elements do not pulsate with amplitudes above 50--100~\ms, which
is at least an order of magnitude lower than the 1--5~\kms\ variability observed in
the lines of rare-earth elements (REE). Many other roAp stars have been found to show a very similar
overall pulsational behaviour \citep[e.g.,][]{KR01b,Balona02,MHK03}.

The peculiar characteristics of the $p$-mode pulsations in roAp stars were clarified by
\citet{RPK02}, who were the first to relate pulsational variability to vertical stratification
of chemical elements. This study of the atmospheric properties of \equ\ showed that
the light and iron-peak elements are enhanced in the lower atmospheric layers, whereas REE ions
are concentrated in a cloud with a lower boundary at $\log\tau_{5000}\la-4$ \citep{MRR05}. Thus,
high-amplitude pulsations observed in REE lines occur in the upper atmosphere while lines of
elements showing no significant variability form in the lower atmosphere. This leads to the
following general picture of roAp pulsations: we observe a signature of a magnetoacoustic wave,
propagating outwards with increasing amplitude through the chemically stratified atmosphere.  The
presence of significant phase shifts between the pulsation radial velocity (RV) curves of different
REEs \citep{KR01a}, or even between lines of the same element \citep*{MHK03}, can be attributed to the
chemical stratification effects and, possibly, to the short vertical wavelength of the running
magnetoacoustic wave. These unique properties of  roAp pulsations, combined with a presence of
large vertical abundance gradients in the line-forming region, make it possible to resolve the
vertical structure of $p$-modes and to study propagation of pulsation waves at the level of detail
previously possible only for the Sun. 

The outstanding pulsational variability of REE lines in the spectra of rapidly rotating roAp stars 
permitted detailed mapping of the horizontal pulsation structure, leading to observational evidence 
that clearly favors one of the two aforementioned theoretical predictions. In fact, the short rotation
periods of some roAp stars and the oblique nature of their non-radial oscillations allow pulsational
monitoring from different aspect angles, thus facilitating reconstruction of the horizontal
pulsation pattern. Using this unique geometrical property of roAp pulsations, and applying pulsation Doppler
imaging and moment analysis techniques, \citet{K04b,K06} has successfully disentangled different
harmonic contributions to the pulsation geometry of the prototype roAp star HD\,83368 and for the
first time obtained an independent confirmation of the alignment of non-radial pulsations and
magnetic field. It was shown that the magnetoacoustic pulsations in HD\,83368 are shaped as
suggested by \citet{SG04}, whereas the non-axisymmetric pulsation components predicted by the theory
of \citet{BD02} cannot be detected.

As for the slowly rotating roAp stars, despite dramatic recent progress in understanding the
vertical structure of their pulsation modes, relatively little attention has been paid to the
problem of inferring the  horizontal geometry of pulsations.
Typically it is assumed that a horizontal cross-section of non-radial pulsation is given by
an oblique axisymmetric mode of low angular degree, similar to the pulsation geometries inferred for rapidly rotating roAp
stars, where periodic modulation of the geometrical aspect supplies additional information that
helps to constrain the pulsation geometry. Thus, the question of systematic mode identification,
central to the studies of other types of pulsating stars, has not been thoroughly investigated in
the case of sharp-lined magnetic pulsators, which represent the majority of roAp stars. 

Understanding rapid line profile variations (LPV) of the slowly rotating roAp stars turns out to
be a challenging task. The first observation that achieved a signal-to-noise ratio sufficient to
detect profile variability \citep{KR01a} demonstrated the presence of an unusual blue-to-red
running feature in the residual spectra of \equ. Moreover, observations of a single-wave
variability of the REE line width in \equ\ is clearly inconsistent
with \textit{any} axisymmetric pulsation geometry described by spherical harmonics 
\citep[see][]{APW92}. This led \citet{KR01a} to attribute LPV in \equ\ to a \textit{non-axisymmetric} $\ell=2$--3
non-radial mode, contrary to the classical OPM of \citet{K82} and in contrast to their own results for rapid
rotators. Later \citet{SKK04} drew attention to the blue-to-red running wave in our time-resolved
spectra of \equ. These authors showed that the observed LPV is inconsistent with spectral
variability expected for any low-degree mode in a slowly rotating star and, in turn, suggested an
exotic shock wave model to explain the observed rapid line profile changes.

In this study we embark on the task of detecting, characterizing and interpreting LPV in sharp-lined
roAp stars. Such a study has become feasible thanks to the availability of a large amount of high-quality,
time-resolved spectroscopic data for a sizable sample of roAp stars. The aim of our analysis is 
to derive from observations a complete picture of LPV in a number of sharp-lined roAp stars and to
tackle the problem of interpretation of the REE line variability. As a result of our roAp  line
profile variability survey, we reveal ubiquitous asymmetric LPV in many magnetic stars. We
suggest a revised OPM that recognizes the influence of stellar pulsations on turbulence in
the upper atmospheric layers of roAp stars, and in this way resolves the apparent discrepancy between the
axisymmetric pulsation picture required by the standard OPM and asymmetric LPV observed in sharp-lined roAp
stars. 

\begin{table*}
\begin{minipage}{135mm}
\caption{Fundamental parameters of program stars.\label{tbl1}}
\begin{tabular}{rlccrccl}
\hline 
HD~~~  & ~~Other         & \te   & \lgg & \vs~~     & \bs & P          & Reference  \\
number & ~~name          & (K)   &      & (\kms)    & (kG)& (min)      &            \\
\hline
  9289 &BW Cet           & 7840  & 4.15 & 10.5~~    & 2.0 &10.52       & \citet{RSK07} \\
 12932 &BN Cet           & 7620  & 4.15 & 3.5~~     & 1.7 &11.65       & \citet{RSK07} \\
 19918 &BT Hyi           & 8110  & 4.34 & 3.0~~     & 1.6 &11.04       & \citet{RSK07} \\
 24712 &DO Eri, HR 1217  & 7250  & 4.30 & 5.6~~     & 3.1 & 6.13       & \citet{RLG97} \\
101065 &Przybylski's star& 6600  & 4.20 & 2.0~~     & 2.3 &12.15       & \citet{CRK00} \\
122970 &PP Vir           & 6930  & 4.10 & 4.5~~     & 2.3 &11.19       & \citet{RSH00} \\
128898 &$\alpha$~Cir     & 7900  & 4.20 & 12.5~~    & 1.5 & 6.80       & \citet{KRW96} \\
134214 &HI Lib           & 7315  & 4.45 & 2.0~~     & 3.1 & 5.69       & \citet{RSK07} \\
137949 &33 Lib           & 7550  & 4.30 &$\le$2.0~~~& 5.0 & 8.27       & \citet{RNW04} \\
201601 &$\gamma$~Equ     & 7700  & 4.20 &$\le$1.0~~~& 4.1 &12.21-12.45 & \citet{RPK02} \\
\hline                               
\end{tabular}                        
\end{minipage}
\end{table*}

The rest of the paper is organized as follows. Sect.~\ref{targets} explains our choice of the
target roAp stars. Sect.~\ref{observ} describes acquisition and reduction of time-series 
spectra. Investigations of spectroscopic variability with residual spectra and moment analysis
are presented in Sect.~\ref{varia}. Interpretation of the observations using theoretical spectrum synthesis is given
in Sect.~\ref{newopm}. Results of our study are summarized and discussed in Sect.~\ref{disc}.

\section{Target selection}
\label{targets}

The pulsational LPV pattern of a roAp star strongly depends on the orientation of the pulsation axis with
respect to the line of sight, and on the magnitude of Doppler shifts due to the stellar
rotation. In the slowly rotating oblique pulsators (\vp\,$\ga$\,\vs), the velocity field observed in
REE lines is dominated by the pulsational perturbations. These roAp stars are
qualitatively similar to normal non-radially pulsating stars, except that the pulsation axis
obliquity $\beta$ plays the role of the inclination angle $i$. When the rotational Doppler shifts 
exceed the local pulsational velocity over a significant fraction of the visible stellar
surface (\vp\,$\ll$\,\vs), the pulsational profile variability pattern changes its form, because it is now determined
by a complex superposition of the rotational velocity field and oblique pulsations \citep{K04a}.
The behaviour of the line profile moments is also modified in comparison to the sharp-lined stars
\citep{K05}. These circumstances suggest that different analysis strategies are best suited for
fast and slow rotators. The former should be observed over the whole rotation period and modelled
taking into account the complete oblique rotator/pulsator geometry. On the other hand, one can, at
least as a first approximation, neglect rotation for the latter stars and treat them as normal
pulsators. The scope of the present paper is observation and interpretation of the pulsational
LPV in the second, slowly rotating, group of roAp stars.

Based on our numerical simulations of the line profile and moment variations \citep{K04a,K05}, we
have established \vs\,$\approx10$--15~\kms\ (the exact figure depends on the pulsation geometry and
amplitude) as an upper limit of the rotational Doppler shifts which can no longer be neglected in
the spectroscopic modelling of oblique non-radial pulsations. This criterion, together with the availability
of high-quality spectroscopic time-series data, has defined the selection of the program stars.
Table~\ref{tbl1} summarizes the main properties of ten roAp stars investigated in our paper. When
available, we give \te, \lgg, \vs\ and the mean field modulus \bs\ derived in recent model
atmosphere and chemical abundance analyses (see references in Table~\ref{tbl1}). For 4 poorly
studied roAp stars (HD\,9289, HD\,12932, HD\,19918, and HD\,134214) no detailed studies have been 
published, hence we adopted atmospheric parameters determined by \citet{RSK07} using Str\"omgren
photometric calibrations \citep{MD85,NSW93}. Estimates of rotational velocity and magnetic field 
for these four stars were obtained by fitting theoretical spectra to unblended spectral lines of
different magnetic sensitivity. Details of this procedure can be found in \citet{RSK07}.

Time-resolved spectra are also available for several roAp stars characterized by moderate to large 
line broadening: HD\,42659, HD\,60435, HD\,80316, HD\,83368, HD\,84041, HD\,99563. These stars do
not satisfy our \vs\ criterion or have observations of low quality (HD\,60435), and were not included
in the analysis.

We note that some objects selected according to their low \vs\ may be relatively rapid
rotators seen close to one of the poles. The 3.877~d rotation period obtained by \citet{RWA05}
for HD\,122970 and the 4.471~d rotational modulation found for HD\,128898 \citep{KSM94,KR01b}
hint on such a viewing geometry. Thus, an additional uncertainty may enter interpretation of
the spectroscopic variability of HD\,122970, HD\,128898 and HD\,9289 (whose rotation period is
unknown, but cannot be long, given a \vs\ of 10.5~\kms.) Nevertheless, spectroscopic observables
studied in our paper are mostly sensitive to the \textit{projected} rotational velocity and
should not be strongly distorted by the special orientation of HD\,122970, HD\,128898 and
similar objects. Consequently, we make no explicit distinction between intrinsically slow rotators
and pole-on sharp-lined stars.

The pulsation periods listed in Table~\ref{tbl1} correspond to variability due to the dominant pulsation
mode at the time of observations. In this study we do not perform a detailed frequency analysis,
and only deal with short (typically $\approx$\,2h-long) spectroscopic time series. Therefore,
possible small inaccuracies in the adopted pulsation periods have no influence on the results
presented below.

\section{Observations and data reduction}
\label{observ}

\begin{table*}
\begin{minipage}{150mm}
\caption{Journal of time-series spectroscopic observations of roAp stars. Superscript
numbers 1--6 mark different data sets of HD\,201601. \label{tbl2}}
\begin{tabular}{rcccccccc}
\hline
HD~~~  & Instrument/       & Start HJD  & End HJD    &Wavelength  & Number of & Exposure & Overhead & Peak \\
number & telescope         & (2450000+) & (2450000+) &region (\AA)& exposures & time (s) & time (s) & SNR  \\
\hline
9289   & UVES/VLT          & 2920.54506 & 2920.62881 & 4960--6990 &111 & 40 & 25 & 90  \\
12932  & UVES/VLT          & 2921.62234 & 2921.70532 & 4960--6990 & 69 & 80 & 25 & 90  \\
19918  & UVES/VLT          & 2921.52607 & 2921.60905 & 4960--6990 & 69 & 80 & 25 & 100 \\
24712  & UVES/VLT          & 3321.65732 & 3321.74421 & 4960--6990 & 92 & 50 & 22 & 300 \\
101065 & UVES/VLT          & 3071.67758 & 3071.76032 & 4960--6990 &111 & 40 & 25 & 180 \\
122970 & UVES/VLT          & 3069.70977 & 3069.79359 & 4960--6990 &111 & 40 & 25 & 160 \\
128898 & UVES/VLT          & 3073.80059 & 3073.88262 & 4960--6990 &265 & 1.5& 25 & 250 \\
134214 & UVES/VLT          & 3070.77571 & 3070.85848 & 4960--6990 &111 & 40 & 25 & 260 \\
137949 & UVES/VLT          & 3071.76312 & 3071.84598 & 4960--6990 &111 & 40 & 25 & 350 \\
201601 & CES/ESO 3.6-m$^1$ & 1381.78344 & 1381.82390 & 6140--6165 & 31 & 60 & 52 & 190 \\
201601 & Gecko/CFHT$^2$    & 2186.70618 & 2186.80296 & 6543--6656 & 64 & 90 & 43 & 250 \\
201601 & Gecko/CFHT$^2$    & 2186.82456 & 2186.92308 & 6619--6729 & 65 & 90 & 43 & 230 \\
201601 & Gecko/CFHT$^3$    & 2540.72237 & 2540.83199 & 5283--5343 & 65 & 90 & 53 & 160 \\
201601 & Gecko/CFHT$^3$    & 2540.86101 & 2540.96843 & 5821--5887 & 65 & 90 & 53 & 130 \\
201601 & Gecko/CFHT$^4$    & 2541.85433 & 2541.93164 & 5821--5887 & 50 & 90 & 43 & 140 \\
201601 & Gecko/CFHT$^5$    & 2542.82688 & 2542.95051 & 6542--6657 & 60 & 90 & 43 & 170 \\
201601 & Gecko/CFHT$^6$    & 2543.82314 & 2543.92191 & 6104--6194 & 64 & 90 & 43 & 230 \\
\hline
\end{tabular}
\end{minipage}
\end{table*}

\subsection{Echelle spectra}

The main spectroscopic data set analysed in our study includes 958 observations of 8 roAp stars obtained
with the UVES instrument at the ESO VLT between October 8, 2003 and March 12, 2004 in the context of
the observing program 072.D-0138 \citep*{KEM06}. The ESO Archive facility was used to search and retrieve science
exposures and the respective calibration frames. For these observations the red arm of the UVES
spectrometer was configured to observe the spectral region 4960--6990~\AA\ (central wavelength
6000~\AA). The wavelength coverage is complete, except for a 100~\AA\ gap centred at 6000~\AA.
Observations were obtained with the high-resolution UVES image slicer (slicer No. 3), providing
improved radial velocity stability and giving a maximum resolving power of
$\lambda/\Delta\lambda\approx115\,000$.

Observations of each target cover 2 hours and consist of an uninterrupted spectroscopic time series
with a total number of exposures ranging from 69 to 265. The length of individual exposures was 40~s or
80~s,  except for the brightest roAp star HD\,128898 ($\alpha$~Cir), for which a 1.5~s exposure time
was used. The ultra-fast (625kHz/4pt) readout mode of the UVES CCDs made it possible to limit overhead to
$\approx$\,25~s, thus giving a duty cycle of 70--80\% for the majority of targets. The signal-to-noise
ratio of individual spectra is between 90 and 350, as estimated from the dispersion of the stellar
fluxes in the line-free regions.  Detailed description of observations for each star is given in
Table~\ref{tbl2}.

The red 600~nm UVES data set was complemented by the observations of HD\,24712 obtained on November 12,
2004 on DDT program 274.D-5011. 92 time-resolved spectra were acquired with the UVES spectrometer,
which was used in the 390+580~nm dichroic mode (wavelength coverage 3300--4420~\AA\ and 4790--6750~\AA).
A detailed description of the acquisition and reduction of these data is given by \citet{RSW06}.

All \'echelle spectra were reduced and normalized to the continuum level with a routine specially
developed by one of us (DL) for fast reduction of spectroscopic time-series observations. A special
modification of the Vienna automatic pipeline for \'echelle spectra processing \citep{tsymbal} was
developed. All bias and flat field images were median averaged before calibration. The scattered light
was subtracted using a 2-D background approximation. For cleaning cosmic ray hits we applied an
algorithm which compares the direct and reversed spectral profiles. To determine the boundaries of
\'echelle orders, the code used a special template for each order position in each row across the 
dispersion axis. The shift of the row spectra relative to the template was derived by a cross-correlation
technique. Wavelength calibration for each roAp star was based on a single ThAr exposure,
recorded immediately after the respective stellar time series. Calibration was done by a 2-D approximation of the
dispersion surface. An internal accuracy of 30--40~\ms\ was achieved by using several hundred ThAr
lines in all \'echelle orders. The final step of continuum normalization and merging of \'echelle orders
was carried out by transformation of the flat field blaze  function to the response function in each
order.

The global continuum normalization was improved by iterative fitting of a smoothing spline function to
the  high points in the average spectrum of each star. With this procedure we have corrected  an
underestimate of the continuum level, unavoidable in the analysis of small spectral regions of the
crowded spectra of cool Ap stars. Correct determination of the absolute  continuum level is important
for retrieving unbiased amplitudes of radial velocity variability and for studying LPV. In addition
to the global continuum correction, the spectroscopic time series were post-processed to ensure
homogeneity in the continuum normalization of individual spectra. Each extracted spectrum was divided by
the mean spectrum. The resultant ratio was heavily smoothed, and was then used to correct the continuum in the individual
spectra. 

\subsection{Single order spectra}

The time-resolved observations of HD\,201601 were obtained in 2001--2002 using the single-order
$f/4$ Gecko coud\'e spectrograph with the EEV1 CCD at the 3.6-m Canada-France-Hawaii telescope. In
total we gathered 491 observations, covering several spectral windows (see Table~\ref{tbl2}) in the
5823--6729~\AA\ wavelength  region. Observations have a resolving power, determined from the widths
of a number of ThAr comparison lines, of about 115\,000. In all observations of HD\,201601 we used
an exposure time of 90~s.  Individual exposures have signal-to-noise ratio between 130 and 230.

The spectra were reduced using standard IRAF tasks. Each stellar, flat and calibration frame had a
mean bias subtracted and was then cleaned of cosmic ray hits and extracted to one dimension.
Extracted stellar spectra were divided by an extracted mean flat field, and the continuum was fit
with a third-order Legendre  polynomial, using the same rejection parameters for all spectra so that
the continuum fit is as uniform as possible. The wavelength scale was established using about 40
lines of a ThAr emission lamp, resulting in an RMS scatter about the adopted pixel-wavelength
polynomial (a sixth-order Legendre polynomial) of about  30~\ms. The wavelength scale was linearly
interpolated between ThAr lamp spectra taken before and after the stellar series. The spectra
were not resampled to a  linear wavelength spacing. 

In this study we also used 31 very high resolution time-resolved observations of HD\,201601
(\equ) analysed by \citet{KR01a}. These data were obtained in 1999 with the Coud\'e Echelle
Spectrograph (CES), fiber-linked to the Cassegrain focus of the ESO 3.6-m telescope. The highest
resolution CES image slicer and the ESO CCD\#38 were used, allowing us to reach a resolving power of 
$\lambda/\Delta\lambda=166\,000$ and record spectra in the 6140--6165~\AA\ wavelength interval. We
refer the reader to \citet{KR01a} for other details of the acquisition and reduction of the CES spectra
of HD\,201601.

Post-processing of the extracted spectra of HD\,201601 was done consistently with the procedure
adopted for UVES observations.

\section{Investigation of spectroscopic variability}
\label{varia}

\subsection{Choice of spectral lines}
\label{lines}

Non-radial pulsations in roAp stars are best observed using lines of the rare-earth ions. Here
we aim at detailed investigation of the pulsational LPV and, therefore, are limited to the
analysis of variability in a few of the strongest REE lines which show the clearest signatures of
oscillations. Our experience, and the results of previous studies, suggest that strong lines of \pr\
and \nd\ are most useful for the detection and interpretation of the rapid LPV. Therefore, we
have chosen a sample of unblended doubly ionized lines of Pr and Nd. In addition, we studied a
few weaker lines of \ndt\ and \tb\ in order to sample different layers in the upper
atmospheres of roAp stars and to resolve possible depth dependence of LPV. The
VALD \citep{VALD} and DREAM \citep{DREAM} databases, complemented with the extended \nd\ line
list \citep{RRK06}, were used as a basis for line identification. In total, we looked at
17 REE lines, including \pr\ 5284, 5300, 6160, 6706~\AA, \ndt\ 5311, 5319, 6650~\AA,
\nd\ 5286, 5294, 5802, 5851, 6145, 6550, 6691~\AA, and \tb\ 5505, 5847, 6323~\AA. Not all lines can
be studied in each star because of blending considerations, different spectral coverage of
the time-resolved spectra, and the intrinsic star-to-star variation in pulsational behaviour.
We preferred to analyse lines with $\lambda\le$\,6000~\AA\ in stars observed with
UVES, whereas red REE lines with central wavelengths above 6000~\AA\ were studied in the CFHT
observations of \equ. In fact, an exhaustive investigation of all the lines listed above in each
star is unnecessary because the information content of many REE features is redundant (this
especially concerns the numerous strong \nd\ lines).

A remarkable diversity of the pulsational amplitudes and phases observed in the spectral lines of
different chemical elements indicate the importance of the depth dependence of chemical abundances
and of pulsation wave properties in roAp atmospheres. Taking this into account, we 
expect to find a significant depth dependence of the pulsational LPV. This is why it is
important to study variability in diverse REE lines, and to associate observed LPV patterns with
specific atmospheric layers. However, the question of the formation depth of REE lines in the
atmospheres of roAp stars is an extremely complex one. Ideally it should be addressed with
detailed NLTE chemical stratification calculations \citep*{MRR05}, possibly taking into account
the anomalous structure of the upper atmospheres of cool Ap stars \citep{KBB02}. Such an analysis is
extremely demanding in terms of computing time and input physics, and up to now has been
carried out only for \ndt\ and \nd\ in two representative roAp stars \citep*{MRR05}.
The results of these elaborate NLTE-chemical stratification calculations show that line intensity can
be used as a good proxy for the \textit{relative} formation depth of spectral features
belonging to the same element. For instance, weaker singly ionized Nd lines are formed deeper
than strong \nd\ lines. On the other hand, the intensity of the lines of different elements
is not a reliable indicator of the formation depth because there is no reason to assume that
vertical stratification is the same for all REE species. Nevertheless, one can still assign an
approximate relative formation depth to the lines of different REEs based on phases of RV
variability. In the framework of a running magnetoacoustic wave which propagates outwards, the 
lines showing later RV maximum are formed higher in the atmosphere. We will use this property
for intercomparison of the results obtained for Pr, Nd and Tb lines. One has to keep in mind,
however, that different horizontal sampling of the pulsation velocity field due to dissimilar
spotted distribution of different REEs may distort this phase-depth relation. Moreover, any
judgment about formation heights based solely on the time of the velocity extrema becomes
meaningless when the vertical $p$-mode structure is dominated by the standing wave component,
and pulsation nodes with the associated 180\degr\ phase change are observed in the line-forming
region. One roAp star included in our sample, HD\,137949, is known to show this behaviour
\citep*{MHK03}.

\subsection{Residual spectra}
\label{resid}

We start analysis of the pulsational LPV in roAp stars by investigating the behaviour of the residual
spectra as a function of pulsation phase. This is the most general approach to characterize LPV of
pulsating stars and, because the whole information content of the line profile changes is 
considered, this method may be preferable in some cases to analysis of average
secondary quantities, such as line profile moments and bisectors. On the other hand, analysis of
the residual spectra has a drawback, in that it is impossible to directly relate features in the LPV
pattern to physical properties of non-radial pulsations. LPV represents a result of a non-trivial
convolution of the intrinsic stellar line profiles and velocities due to pulsation and stellar
rotation. Therefore, credible interpretation of the residual profile variation is possible
only with the help of spectrum synthesis calculations which adequately describe the aforementioned
effects\footnote{In this context we note that the assertion by \citet{SKK04} that certain features of the LPV
observed in \equ\ by \citet{KR01a} can be interpreted in terms of the pulsation velocity amplitude is
erroneous.}.

Investigation of the residual spectra is feasible only for time-resolved observations of
sufficient quality. The signal-to-noise ratio of the time-series spectra of roAp stars available to
us is quite uneven, which precludes direct detection of the periodic LPV in fainter stars (about
half of our sample). To circumvent this difficulty, we developed a special spectral coaddition
procedure. First, we derive an average spectrum for each star and subtract it from the individual
observations. Then pulsation phases are computed using the periods listed in Table~\ref{tbl1}.
Observations with similar phases are averaged. We typically divided the pulsation cycle into 10 phase 
bins, except for the CFHT and ESO 3.6-m \equ\ data sets where 11 and 13 phase bins were used,
respectively.

\begin{figure}
\fifps{7.0cm}{90}{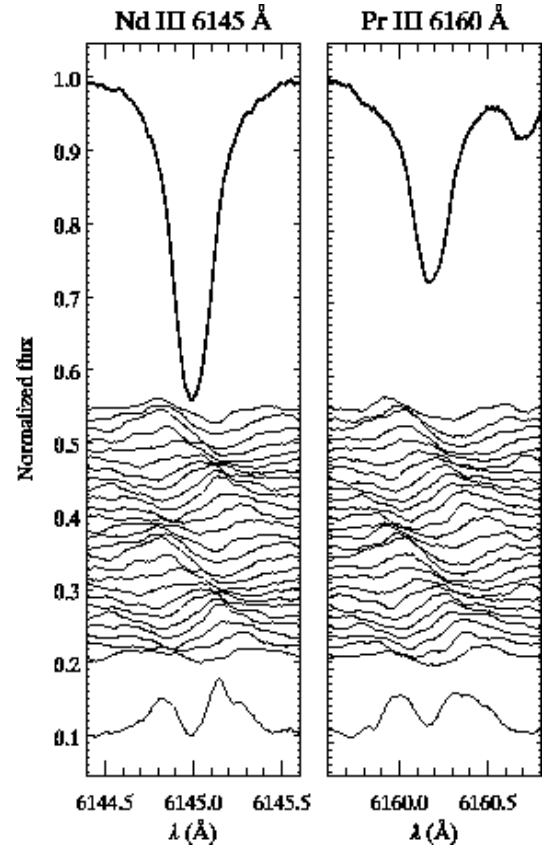}
\caption{Profile variations of the \nd\ 6145~\AA\ and \pr\ 6160~\AA\ lines in the CES
observations of \equ\ obtained on 22 June, 1999. The average spectrum is plotted in the upper part of each panel.
The difference spectra as a function of pulsation phase are presented in the middle part. 
Observations for consecutive pulsation
phases are shifted in the vertical direction, and each phase is shown twice. The bottom
curve presents the wavelength dependence of the the standard deviation (expanded by a factor of 10).} 
\label{fig1}
\end{figure}

Variation of the residual profiles of the \nd\ 6145~\AA\ and \pr\ 6160~\AA\ lines in \equ,
processed with our phase binning procedure, is illustrated in Fig.~\ref{fig1}. In this and other
similar plots we show the average line profile on top, the time series of the difference spectra is
plotted in the middle (with time increasing downward), and the standard deviation profile is shown at the bottom of the plot. Comparison
of the binned LPV presented in Fig.~\ref{fig1} with the profile variations in the original observations
\citep[see fig.~1 in][]{KR01a} shows very good agreement. This indicates that application of the
phase binning procedure to the time-resolved spectra of roAp stars is able to enhance $S/N$ while
retaining the original LPV pattern. As was observed by \citet{KR01a}, we find that LPV of
the \nd\ and \pr\ lines studied is asymmetric, and is dominated by the blue-to-red moving feature.

\begin{figure*}
\fifps{15cm}{90}{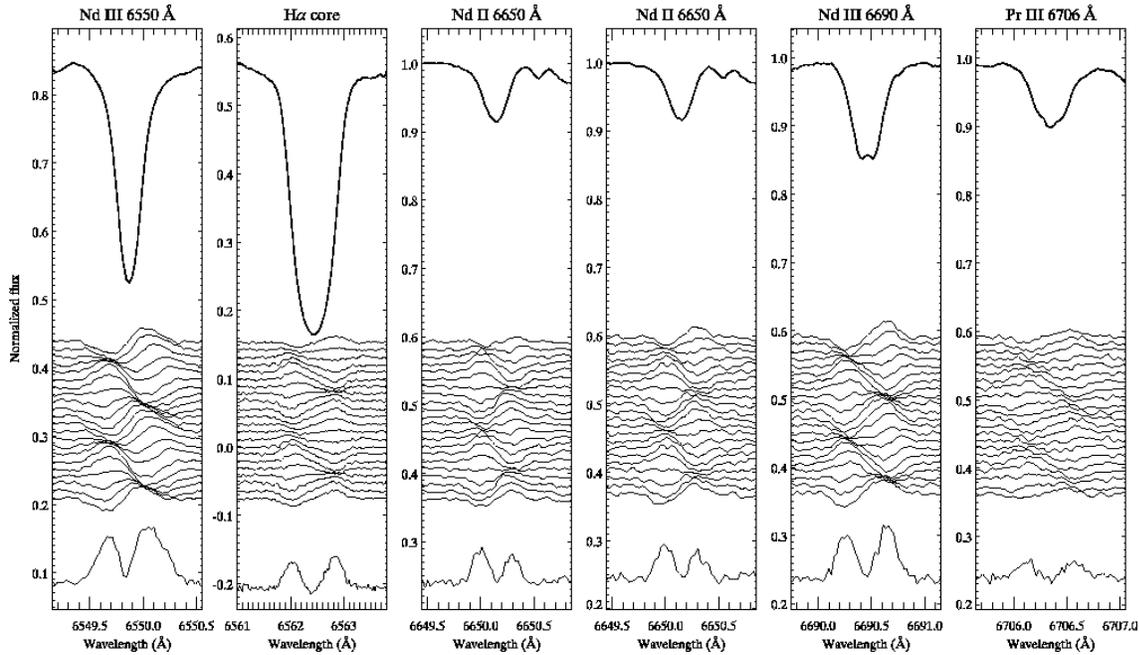}
\caption{Profile variations of the \nd\ 6550, H$\alpha$ core, \ndt\ 6650, \nd\ 6691, and
\pr\ 6707~\AA\ lines in the CFHT observations of \equ\ obtained on 4 October, 2001. The layout
of the figure is similar to Fig.~\ref{fig1}.} 
\label{fig2}
\end{figure*}

The LPV pattern detected in the CFHT observations of \equ\ is shown in Fig.~\ref{fig2}. This \equ\
data set is unique in that it shows an exceptionally high amplitude of the pulsational variability (RV
semi-amplitude of up to 1~\kms\ for a number of REE lines) compared to all other observations of
\equ\ reported in the literature \citep{KR01a,KRP04,SHM05}. Fig.~\ref{fig2} compares behaviour of
the core of \ha, \pr\ 6706~\AA, \nd\ 6550, 6691~\AA\ and \ndt\ 6650~\AA. (Two plots are shown for the latter line,
corresponding to observations in the two consecutive time series recorded during
the same night.) One can see a clear difference in the LPV pattern displayed by the \ha\ core,
\ndt\ 6650~\AA, and \nd\ 6691~\AA\ on the one hand and by the \nd\ 6550~\AA, \pr\ 6706~\AA\ features
on the other hand. The first group of lines shows nearly symmetric, S-shaped variability, whereas
the two lines from the second group behave similarly to the \nd\ 6145~\AA\ and \pr\ 6160~\AA\ lines
in Fig.~\ref{fig1}, and show a blue-to-red moving wave. We suggest that this difference reflects
the depth variation of the pulsational LPV pattern. Sophisticated line formation calculations
\citep*{MRR05} predict that the absorption in the \ndt\ 6650~\AA\ line originates at the bottom of
the Nd-rich layer. The \nd\ 6691~\AA\ line is one of the few doubly ionized Nd lines whose
formation depth overlaps with that of \ndt. This is confirmed by a rather small delay
($\Delta\varphi=0.08$, hereafter we give phases and phase differences in units of pulsation
period) of its RV maximum relative to the radial velocity variation of \ndt\ 6650~\AA.
In contrast, the strong line \nd\ 6550~\AA\ forms substantially higher and shows a large phase delay 
($\Delta\varphi=0.21$). Detailed NLTE radiative transfer calculations are not yet available for
\pr, but, given that the weak \pr\ 6706~\AA\ line has a phase lag similar to that of \nd, its
effective formation height should not differ much. Indeed, we see qualitatively the same LPV
pattern for this \pr\ feature and for \nd\ 6550~\AA. 

The formation depth of the \ha\ core is difficult to estimate. Based on its symmetric LPV pattern, we
expect the \ha\ core to form below the Nd-rich layer, in agreement with the proposed depth dependence
of LPV in metal lines. Nevertheless, the large intrinsic Doppler width of the \ha\ core does not permit
a straightforward comparison with the variability detected in REE lines. In addition, the pulsational
changes of the \ha\ core are too low in some of the roAp stars included in our study. For these reasons
we refrain from detailed analysis of the \ha\ profile behaviour in this paper.

The CFHT observations of \equ\ discussed in this section constitute the first detection of the
depth variation of the pulsational LPV pattern in a roAp star. Moreover, to our
knowledge, such behaviour has not previously been observed in any main sequence non-radially
pulsating star. We conclude that the pulsational LPV changes significantly within the REE-enriched
cloud in the upper atmosphere of \equ: the symmetric, S-shaped LPV pattern of the lines from deeper
layers is replaced by asymmetric, blue-to-red moving features in the LPV of lines formed higher up.

For the analysis of the REE line profile variability in our main UVES data set of nine roAp stars we
used the \ndt\ 5319~\AA, \nd\ 5294~\AA, \pr\ 5300~\AA\ and \tb\ 5847~\AA\ lines. The \ndt\ feature
is sufficiently strong to be observed in all stars and is formed at the bottom of the Nd-rich cloud.
The 5294~\AA\ line is one of the strongest \nd\ lines in the optical spectra of cool Ap stars. This
line samples high atmospheric layers. In general, nothing can be said about the formation heights of
the chosen lines of doubly ionized Pr and Tb because they depend on currently unknown
details of stratification of these REE species in individual stars.

\begin{figure*}
\fifps{15cm}{0}{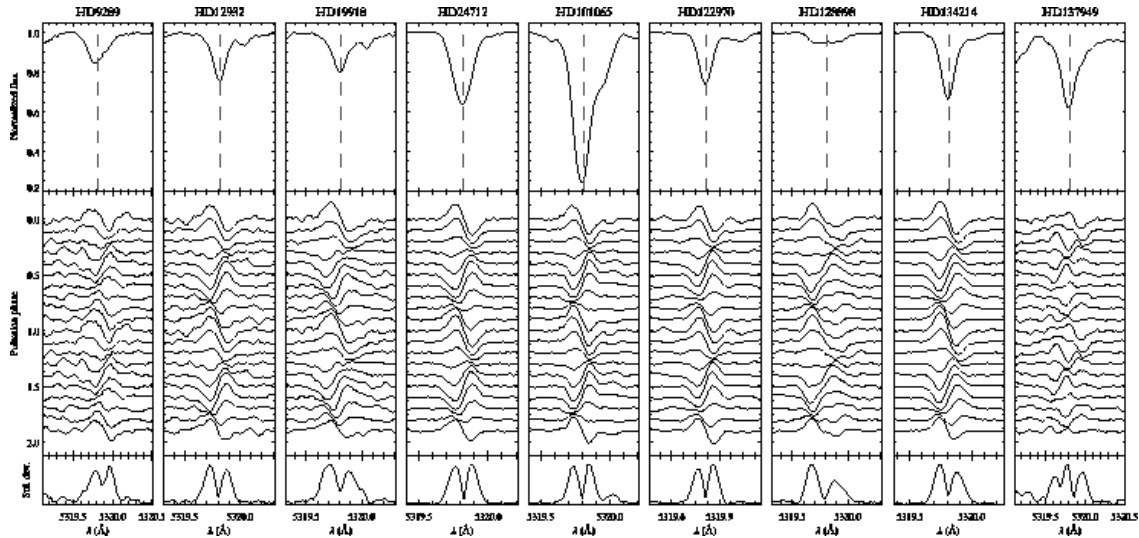}
\caption{Profile variations of the \ndt\ 5319~\AA\ line for nine roAp stars. The layout
of the figure is similar to Fig.~\ref{fig1}, except that here an arbitrary intensity scale is used 
for the residual and standard deviation spectra. The vertical dashed lines indicate position
of line centres.} 
\label{fig3}
\end{figure*}

Fig.~\ref{fig3} shows residual profile variation for the \ndt\ 5319~\AA\ line. Among the nine roAp
stars presented in this plot, seven show a symmetric or nearly symmetric (HD\,19918), S-shaped LPV
pattern similar to the one which we found for \ndt\ lines in \equ. HD\,9289 shows a weak
red-to-blue moving wave -- an odd behaviour which is not seen in any other star. Time series of
the residual spectra are inconclusive for HD\,137949 because at the heights sampled by \ndt\
5319~\AA, a  pulsation node is observed in this star and the variation is dominated by the first
harmonic of the main frequency. However, the \ha\ core in HD\,137949 has symmetric variability
\citep{KEM04}, quite similar to the behaviour of \ndt\ in the majority of the roAp stars studied.

\begin{figure*}
\fifps{15cm}{0}{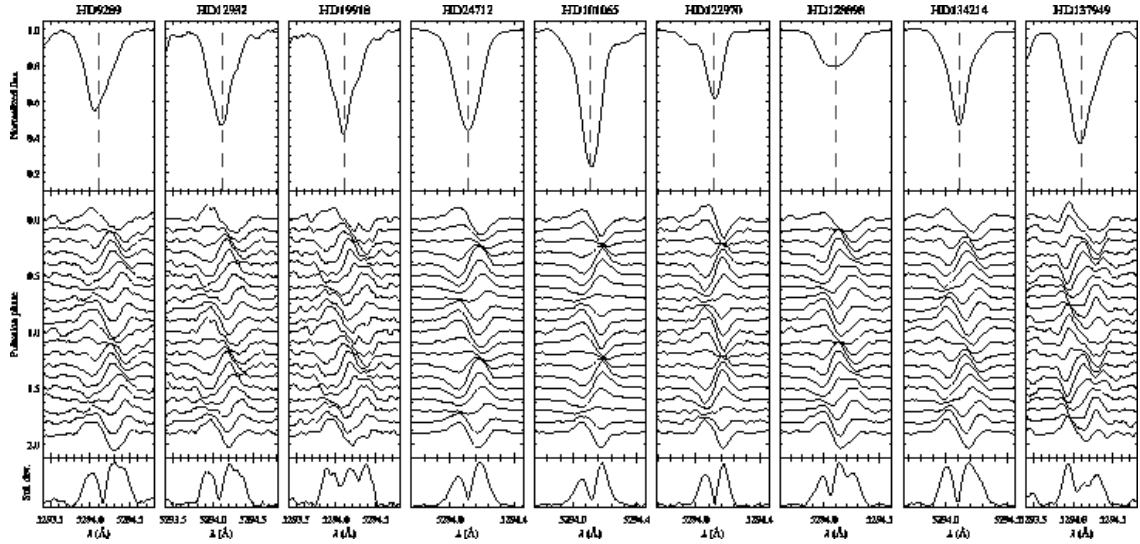}
\caption{Profile variations of the \nd\ 5294~\AA\ line for nine roAp stars. The layout
of the figure is similar to Fig.~\ref{fig3}.} 
\label{fig4}
\end{figure*}

Considering variation of the \nd\ 5294~\AA\ line (Fig.~\ref{fig4}), we detect a well-developed LPV
pattern with blue-to-red moving waves in HD\,9289, HD\,12932, HD\,19918, HD\,24712, HD\,128898,
HD\,134214 and HD\,137949. The phase lag of this \nd\ line with respect to the RV changes of \ndt\
5319~\AA\ is typically in the range 0.17--0.22, but smaller (0.15--0.17) for HD\,24712 and
HD\,134214. These two stars exhibit LPV of a transitional type: it is possible to trace a blue-to-red
moving wave; however, the latter is not continuous, but formed by two separate `bumps' in the
time-sequence of the residual spectra. The two cooler stars, HD\,101065 and HD\,122970, retain the
S-shaped LPV which was present in the \ndt\ 5319~\AA\ line. Remarkably, these two stars also show a
much smaller phase lag ($\Delta\varphi$\,=\,0.01--0.04) between \nd\ and \ndt\ compared to the rest of
the roAp star sample. Either the formation heights of the \ndt\ and \nd\ lines are less different in cooler roAp
stars due to a special form of Nd stratification, or the pulsation wave structure sampled by the Nd
lines is dominated by a standing wave. It is worth mentioning that abundance analysis of HD\,101065
\citep{CRK00} provides no evidence of the \ndt-\nd\ abundance anomaly found in other roAp stars.

\begin{figure*}
\fifps{15cm}{0}{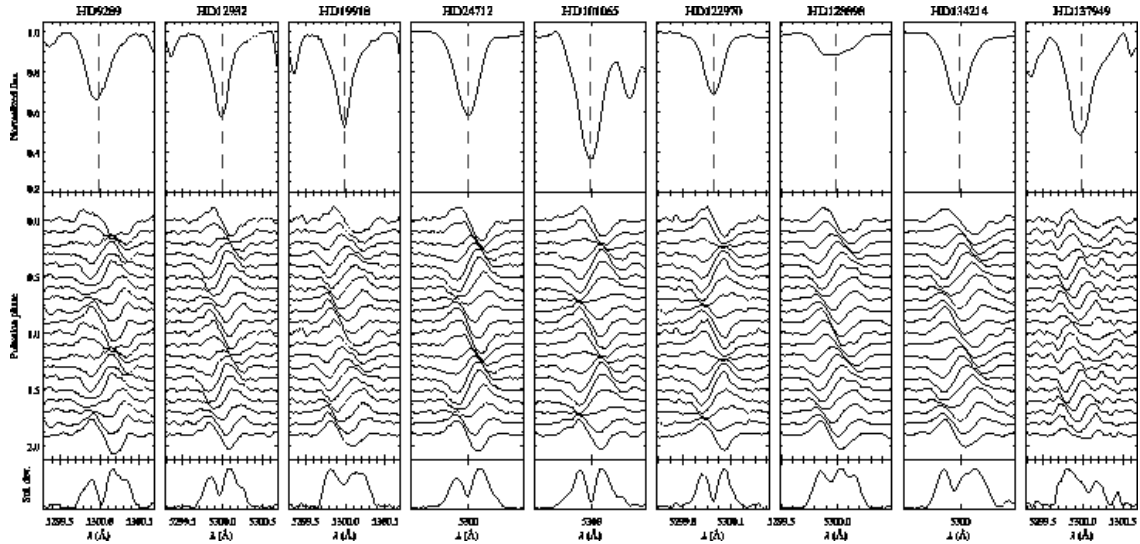}
\caption{Profile variations of the \pr\ 5300~\AA\ line for nine roAp stars. The layout
of the figure is similar to Fig.~\ref{fig3}.} 
\label{fig5}
\end{figure*}

The variation of the \pr\ 5300~\AA\ line is illustrated in Fig.~\ref{fig5}. The negligible phase lag of
this REE line with respect to \nd\ 5294~\AA\ in HD\,9289, HD\,12932 and HD\,19918 corresponds to
essentially the same LPV behaviour as shown by the \nd\ line. Thus, we may suggest that \pr\ and
\nd\ lines in these stars form in similar layers -- a situation reminiscent of the \pr/\nd\ line
variation in \equ. HD\,24712, HD\,128898 and HD\,134214 show a larger phase lag
($\Delta\varphi$\,=\,0.06--0.11) and a stronger asymmetry in the \pr\ line variability. HD\,101065 and
HD\,122970 continue to show symmetric LPV pattern. Analysis of the \pr\ 5300~\AA\ line in
HD\,137949 is complicated by the low RV amplitude and by the presence of harmonic oscillations.
Nevertheless, blue-to-red moving features are still visible in the time evolution of the residual spectra
constructed for this star.

\begin{figure}
\figps{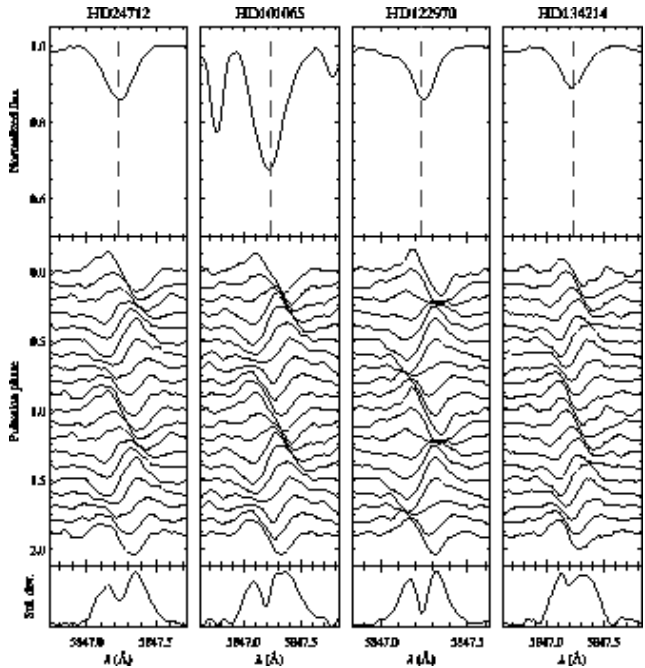}
\caption{Profile variations of the \tb\ 5847~\AA\ line for four cool roAp stars. The layout
of the figure is similar to Fig.~\ref{fig3}.} 
\label{fig6}
\end{figure}

Finally, we investigated variation of the \tb\ 5847~\AA\ line. Its pulsational behaviour is diverse in
the studied sample of roAp stars. Large pulsational amplitude and significant phase shifts of this line
in some stars may indicate that \tb\ probes atmospheric layers above those to which \nd\ and \pr\ lines
are sensitive. Fig.~\ref{fig6} shows variation of \tb\ 5847~\AA\ in four cooler roAp stars (HD\,24712,
HD\,101065, HD\,122970, HD\,134214). A very clear asymmetric LPV and significant (0.16--0.23) phase
delay with respect to \pr\ 5300~\AA\ is found for HD\,24712, HD\,101065 and HD\,134214. Variation of the
\tb\ line in HD\,122970 exhibits a transitional behaviour (cf. the \nd\ 5294~\AA\ variation in HD\,134214),
showing the first signature of the blue-to-red running waves.

Our survey of the pulsational LPV in Nd, Pr and Tb lines in nine roAp stars reveals ubiquitous
signatures of the depth variation of the profile variability pattern in every star. This
corroborates results obtained for \equ\ and shows that a large vertical gradient of the amplitude
and phase of pulsation waves observed in the atmospheres of roAp stars is associated with a similar
rapid change of the LPV pattern. All stars, except HD\,9289, show symmetric LPV in deeper layers. All
stars, without exception, ultimately develop asymmetric LPV in the upper atmospheric region. This
behaviour is apparently correlated with \te: we find that hotter stars (HD\,9289, HD\,12932,
HD\,19918, HD\,128898, HD\,137949) develop interesting asymmetric LPV pattern lower (in the layers
sampled by \nd\ 5294~\AA) than cooler stars (HD\,24712, HD\,101065, HD\,122970, HD\,134214). We
also identified a characteristic change in the shape of the standard deviation profile associated with
the transition from symmetric to asymmetric LPV. In the former case the blue and red peaks in the
standard deviation spectra are well separated and there is almost no variation in the line core. In
the latter case, the two peaks are often merged together and some residual intensity changes
are observed in the line centre (e.g., \pr\ 5300~\AA\ in HD\,19918 and HD\,134214, \tb\ 5847~\AA\ in
HD\,24712 and HD\,134214).

\subsection{Line profile moments}
\label{sect_moments}

Interpretation of the line profile moment variation of slowly rotating pulsating stars is a useful
pulsation mode diagnostic technique \citep*{Balona86,APW92,BA03}. Many studies have successfully utilized
analysis of the pulsational behaviour of the low-order line profile moments to determine
oscillation frequencies and to identify pulsation modes. Recently \citet{K05} has generalized the moment
method to treat pulsations arbitrarily inclined with respect to the stellar rotation axis. This
modification of the moment technique is essential for its application to roAp stars.

In the present study we complement direct analysis of the pulsational LPV based on residual spectra
with the study of pulsational perturbation of the REE line profile moments. These quantities are
defined according to the following relation \citep*{APW92}
\beq
\langle V^j \rangle = 
\sum_i \left(\dfrac{c(\lambda_i-\langle \lambda \rangle)}{\langle \lambda \rangle}\right)^j (1-S_i)/\sum_i (1-S_i),
\label{moments}
\eeq
where $\langle V^j \rangle$ is the line profile moment of $j$th order, $\lambda_i$ and $S_i$ are
the wavelength and normalized intensity at the $i$th pixel of observed spectrum and $c$ is the
speed of light. Moments are computed with respect to the wavelength $\langle \lambda \rangle$
corresponding to the centre-of-gravity of absorption feature in the average spectrum. The first
moment \va\ represents a measure of radial velocity, whereas the second moment \vb\ is related to line
width. The sum in Eq.~(\ref{moments}) is cut off when intensity reaches continuum level or
at the point when blending by the neighboring lines becomes significant.
To improve the accuracy of the calculation of moments with Eq.~(\ref{moments}), we have resampled
the spectra of roAp stars into a fine wavelength grid with the help of cubic splines. In each star
the pulsational variation of the first and second moments was investigated for a set of unblended lines
of \pr, \ndt, \nd\ and \tb.

Time series of the line profile moments are analysed with the least-squares fitting method in order
to characterize variation at the principal frequency and its first harmonic, and to extract the
respective pulsation phases. Moment variation is approximated by the expression
\beq
\langle V^j \rangle = V_0 (t-t_0) + \sum_{i=1}^2 V_{ji} \cos{\{2 \pi [i (t-t_0)/P - \varphi_{ji}]\}}.
\label{cosfit}
\eeq
Here the first term takes into account possible drift of the spectrograph's zero point. The HJD of the
first exposure of the star at a given night (see Table~\ref{tbl2}) was chosen as the reference time
$t_0$. $V_{ji}$ and $\varphi_{ji}$ are, respectively, the amplitude and phase of the $j$th moment
variation with the main period ($i=1$) and half the main period ($i=2$). With the minus sign in 
front of $\varphi_{ji}$ a larger phase corresponds to a later time of maximum of \va\ or \vb. This
phase convention is natural when discussing effects of the outward propagation of pulsation waves 
in the atmospheres of roAp stars. The pulsations periods in Eq.~(\ref{cosfit}) are fixed to the values
listed in Table~\ref{tbl1}.

From theoretical considerations \citep*{APW92,K05} it is expected that the RV of the star exhibiting
low-amplitude linear non-radial pulsations changes sinusoidally, whereas variation of the second
moment may contain a significant contribution of the first harmonic. The amplitude ratio
$V_{21}/V_{22}$ characterizing the overall shape of the \vb\ phase curve is very sensitive to
the horizontal structure of the pulsation velocity field. However, possible non-linearities may lead
to a non-zero harmonic contribution to the first moment variation and could distort the second
moment amplitude ratio diagnostics. Thus, in order to assess possible contribution of the intrinsic
harmonic variability, we fitted the same expression given by Eq.~(\ref{cosfit}) to the time series
of both \va\ and \vb.

\begin{figure*}
\fifps{8.6cm}{0}{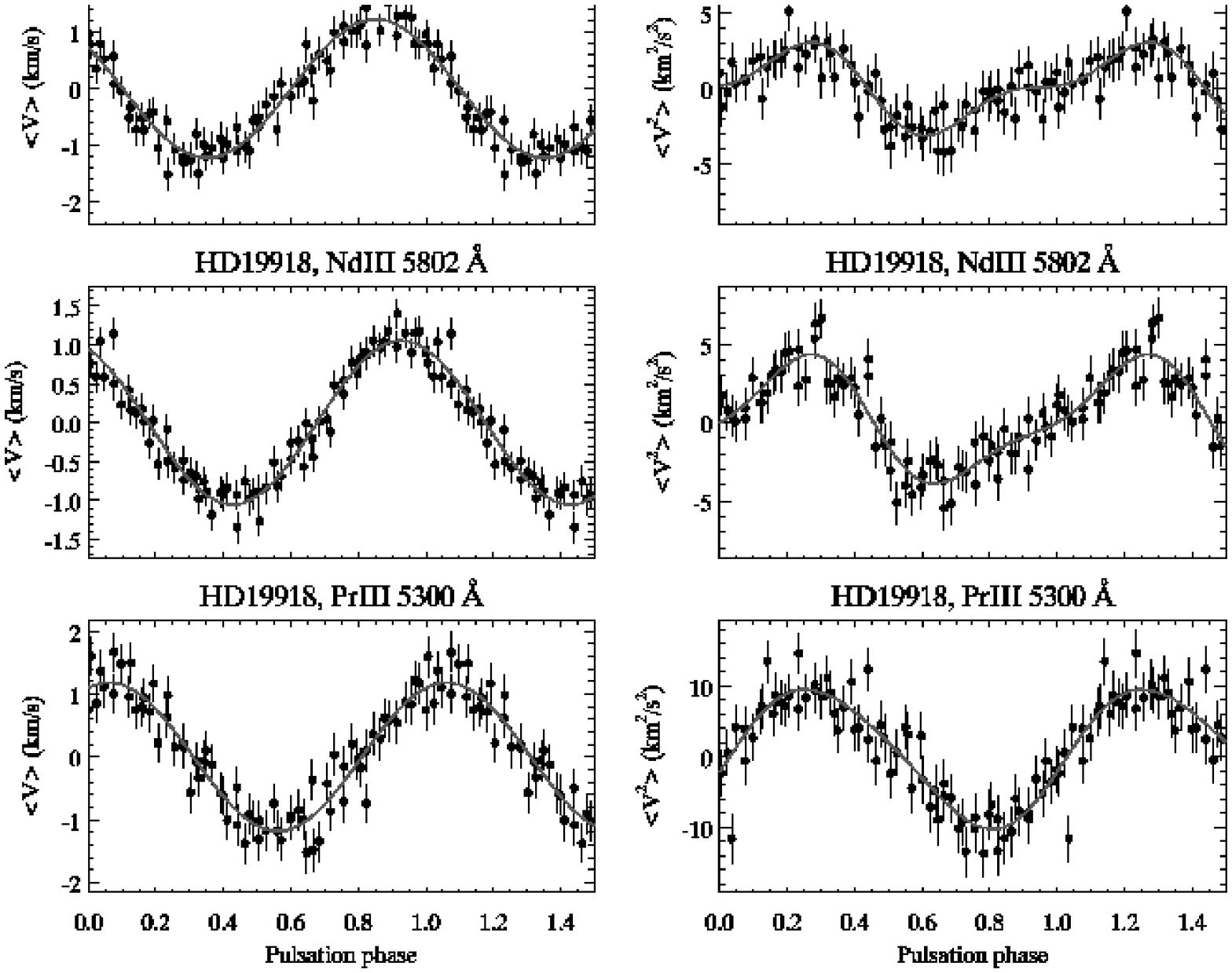}
\fifps{8.6cm}{0}{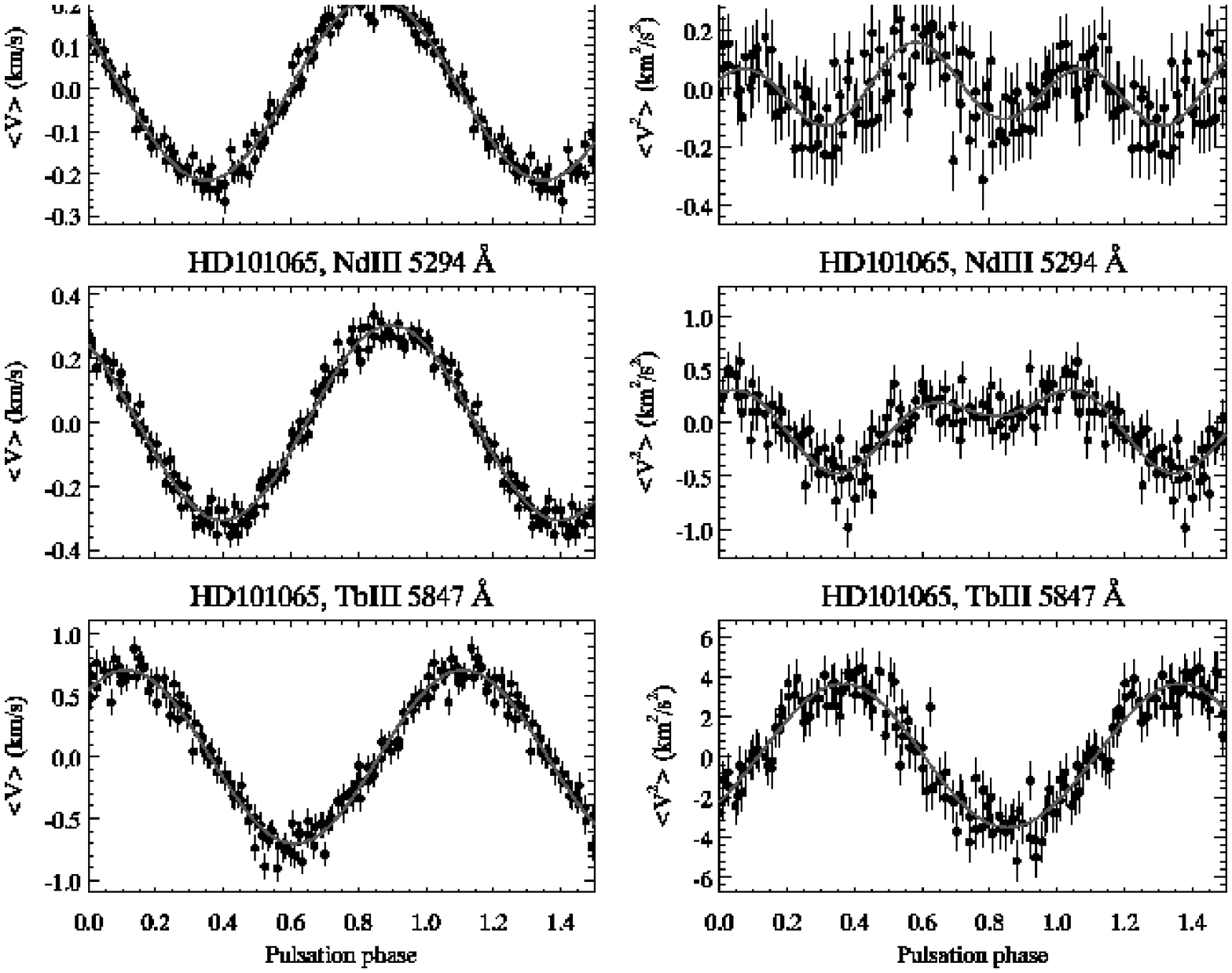}
\caption{Variation of line profile moments for rare-earth lines in {\bf a)} HD\,19918 and 
{\bf b)} HD\,101065. For each star the behaviour of the first \va\ (left panels) and second 
\vb\ (right panels) profile moments is presented. Symbols show the moment measurements as a function
of pulsation phase. The curves illustrate a fit including variation with the principal frequency 
and its first harmonic.}
\label{fig7}
\end{figure*}

An example of the moment analysis for several REE lines in HD\,19918 and HD\,101065 is displayed in
Fig.~\ref{fig7}. The first and second moment measurements are compared with the best-fitting cosine
curves computed according to Eq.~(\ref{cosfit}). Numerical results of fitting the first and second
moment variability are compiled in Tables~\ref{tbl3} and \ref{tbl4} for the roAp stars observed with
UVES and in Table~\ref{tbl5} for HD\,201601. We have verified that the pulsational behaviour of the latter star
does not change significantly in the course of individual observing nights. Thus, we consider moment
measurements obtained in different wavelength regions observed on the same nights to be part of the same
data set. This gives us six groups of  time-resolved spectra, referred to by the numbers 1--6 in
Tables~\ref{tbl2} and \ref{tbl5}.

In the remaining part of this section we explore various trends in the amplitudes and phases of moment
variation and relate these quantities to the LPV of REE lines discussed in Sect.~\ref{resid}. We start
by noting that majority of the roAp stars in our sample show nearly sinusoidal RV variation. Marginally
significant non-zero $V_{12}$ amplitudes are obtained only for a few REE lines in HD\,9289, HD\,101065,
HD\,122970 and HD\,128898. Nevertheless, for all these objects condition $V_{11}\gg V_{12}$ is
fulfilled, indicating that possible non-linearity or a contribution arising from the resonantly excited
harmonic oscillation is negligible. Thus, the presence of harmonic variation of the second moment and
the amplitude ratio $V_{21}/V_{22}$ retains their value as mode structure diagnostics. An atypical
behaviour is found for HD\,137949. A sizable fraction of REE lines in this star shows significant, in
some cases dominant, double-wave RV variation and correspondingly have large $V_{12}$ amplitude. In this
situation, the origin of harmonic variability of \vb\ is ambiguous.

\begin{figure}
\figps{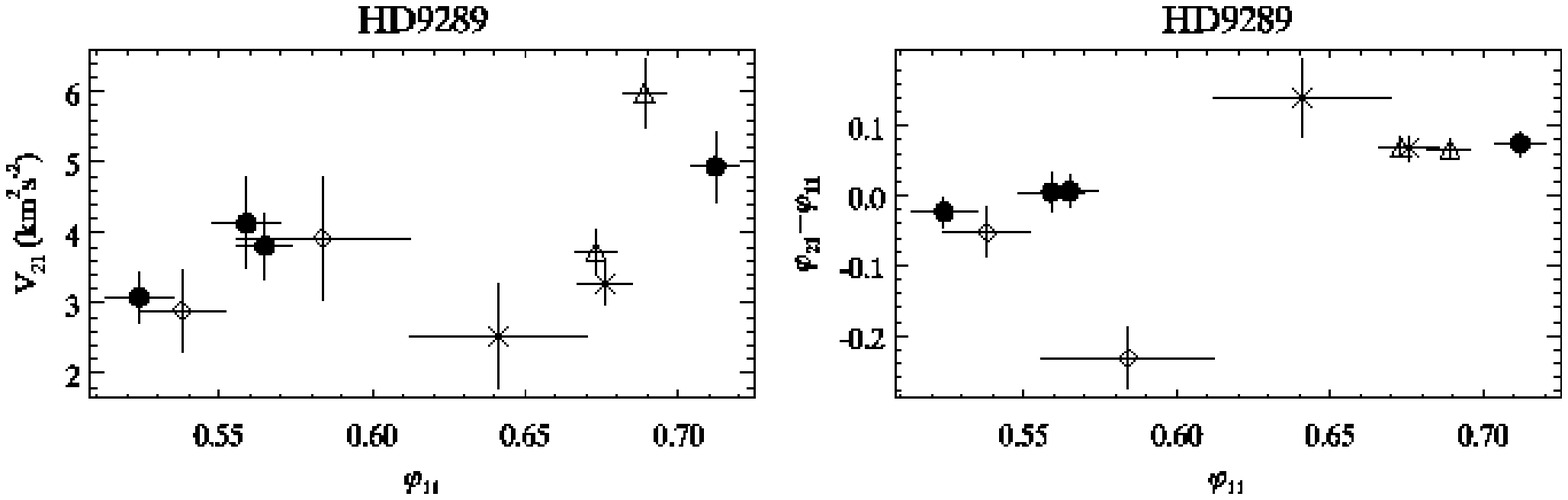}
\figps{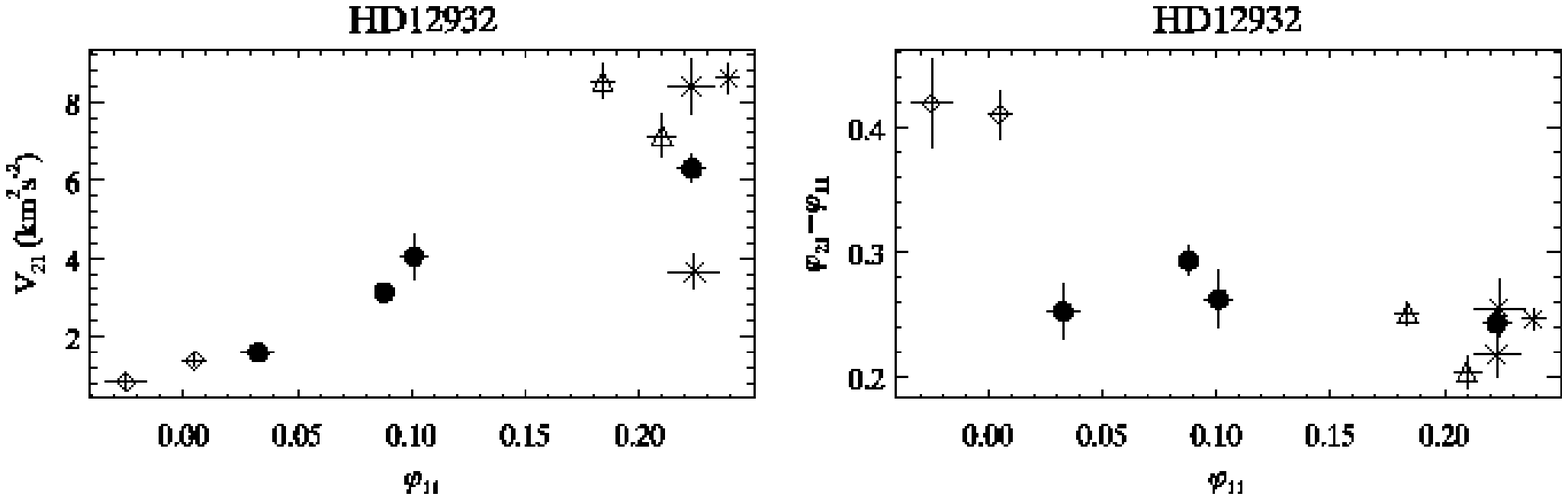}
\figps{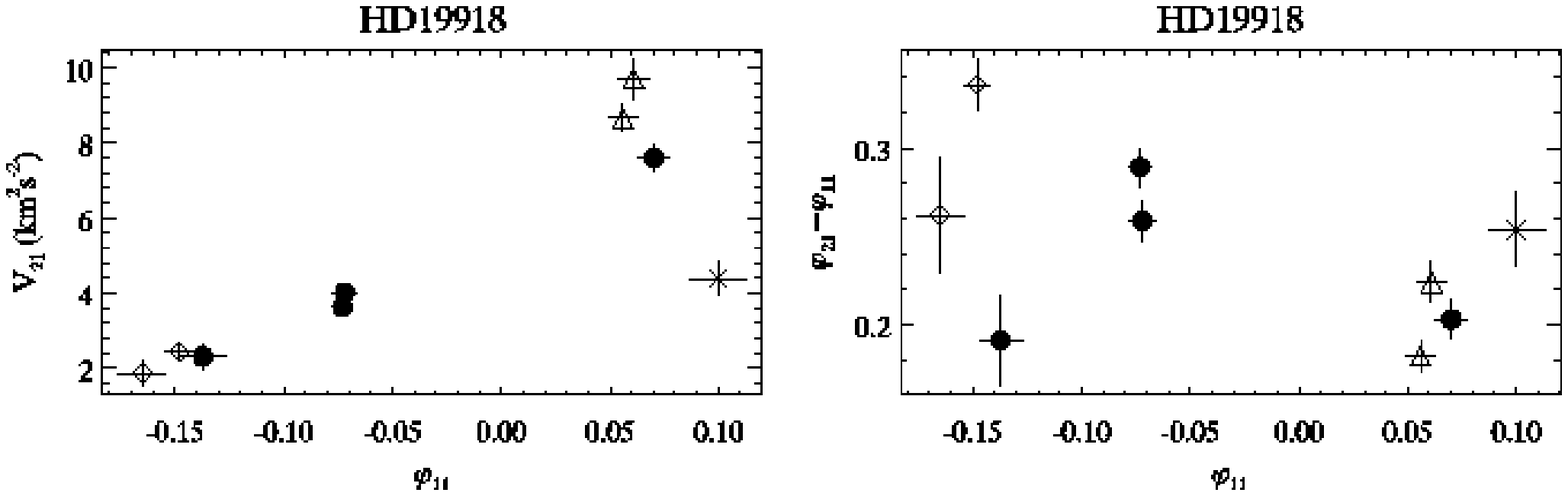}
\figps{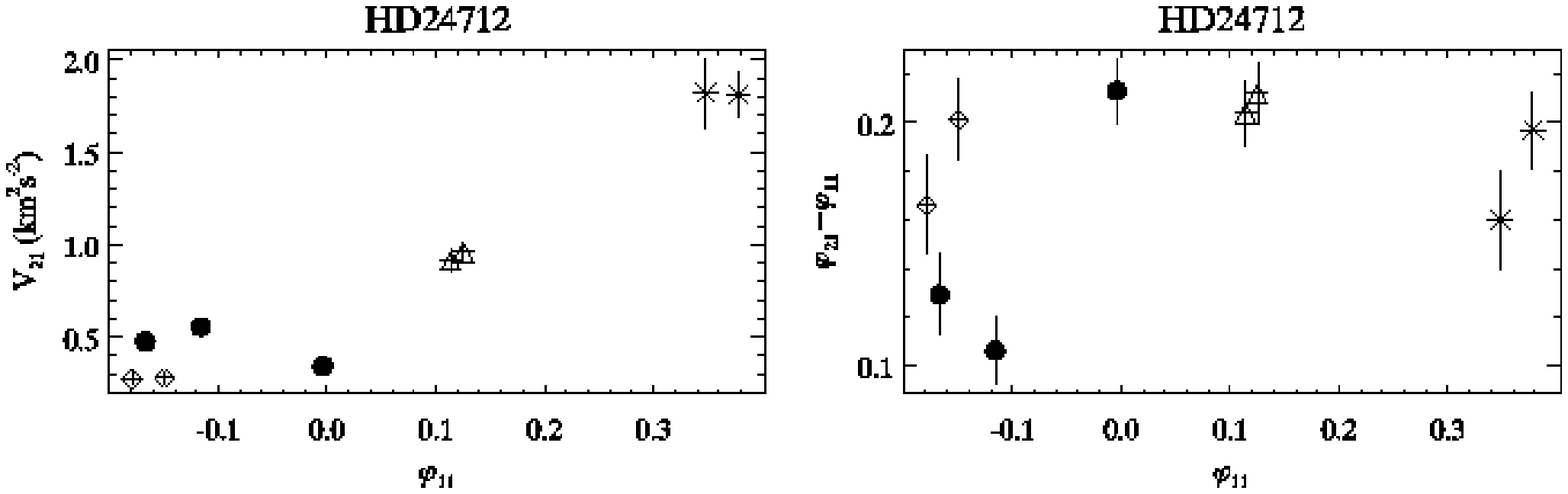}
\figps{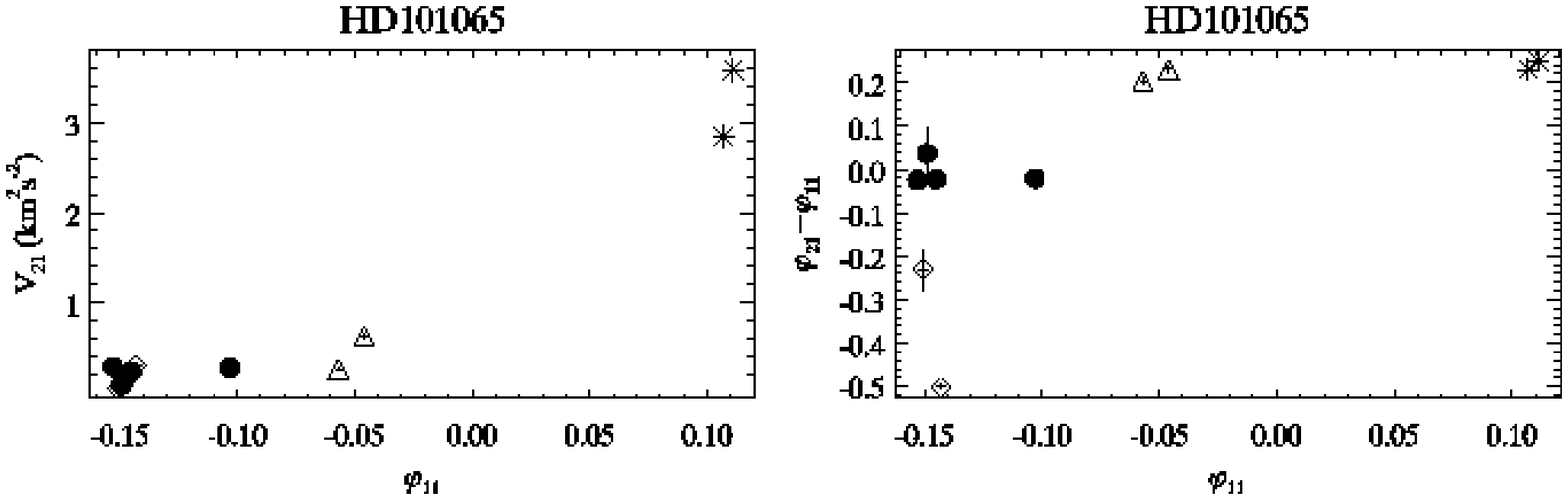}
\caption{The amplitude $V_{21}$ of the second moment variation at the main frequency 
(left panels) and the phase difference $\varphi_{21}-\varphi_{11}$ (right panels) as a 
function of the pulsation phase $\varphi_{11}$ of RV changes for HD\,9289, HD\,12932, HD\,19918,
HD\,24712 and HD\,101065. Symbols correspond to different REE ions: \pr\ ($\triangle$), 
\ndt\ ({\Large$\diamond$}), \nd\ ({\large$\bullet$}), \tb\ ({\Large$\ast$}).}
\label{fig8}
\end{figure}

\begin{figure}
\figps{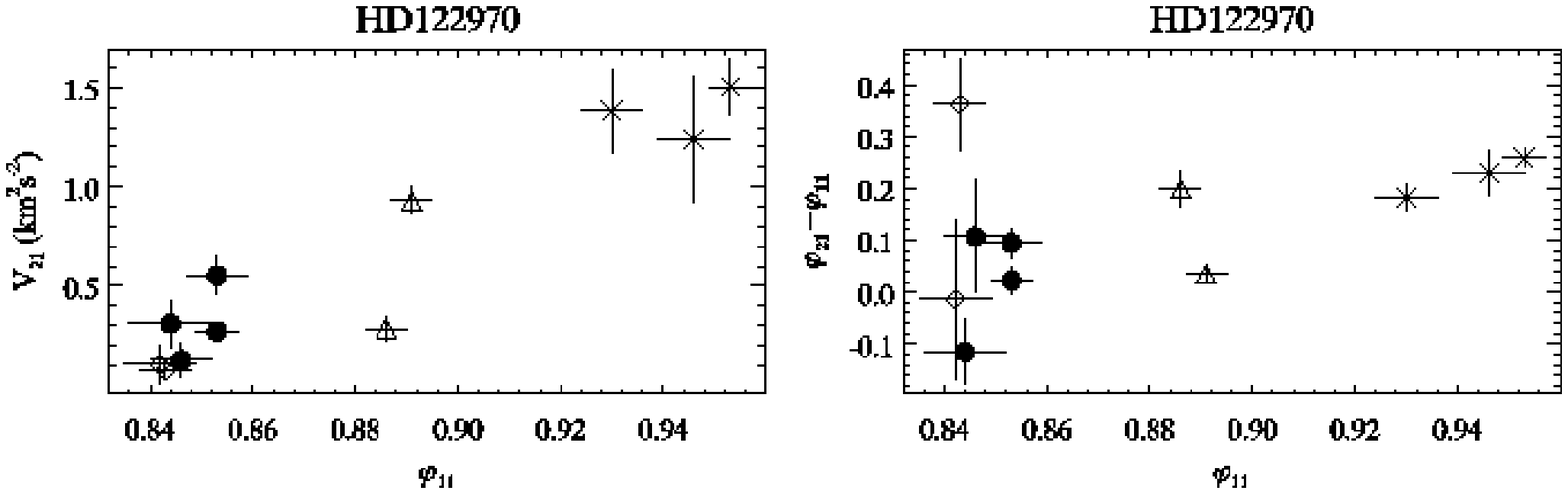}
\figps{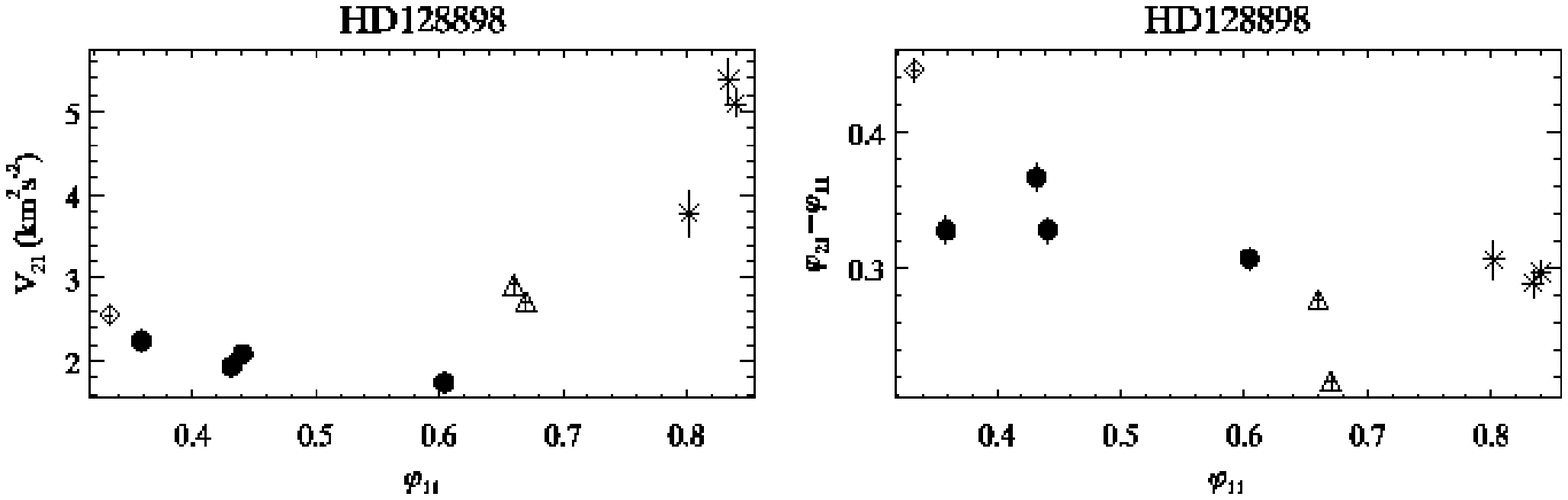}
\figps{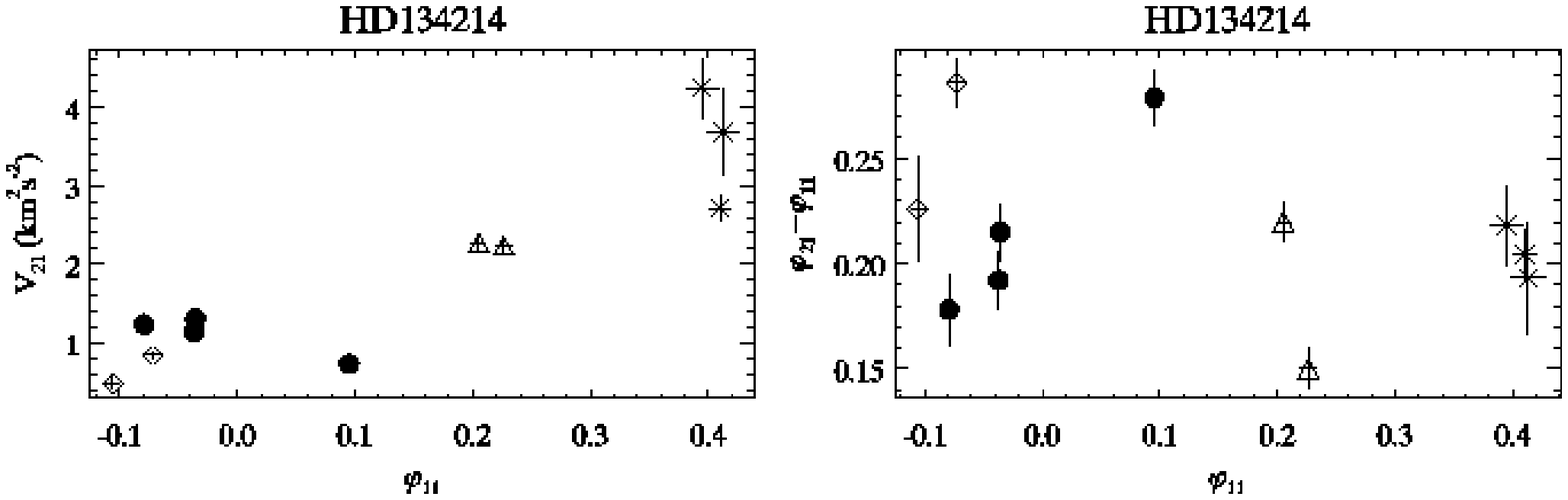}
\figps{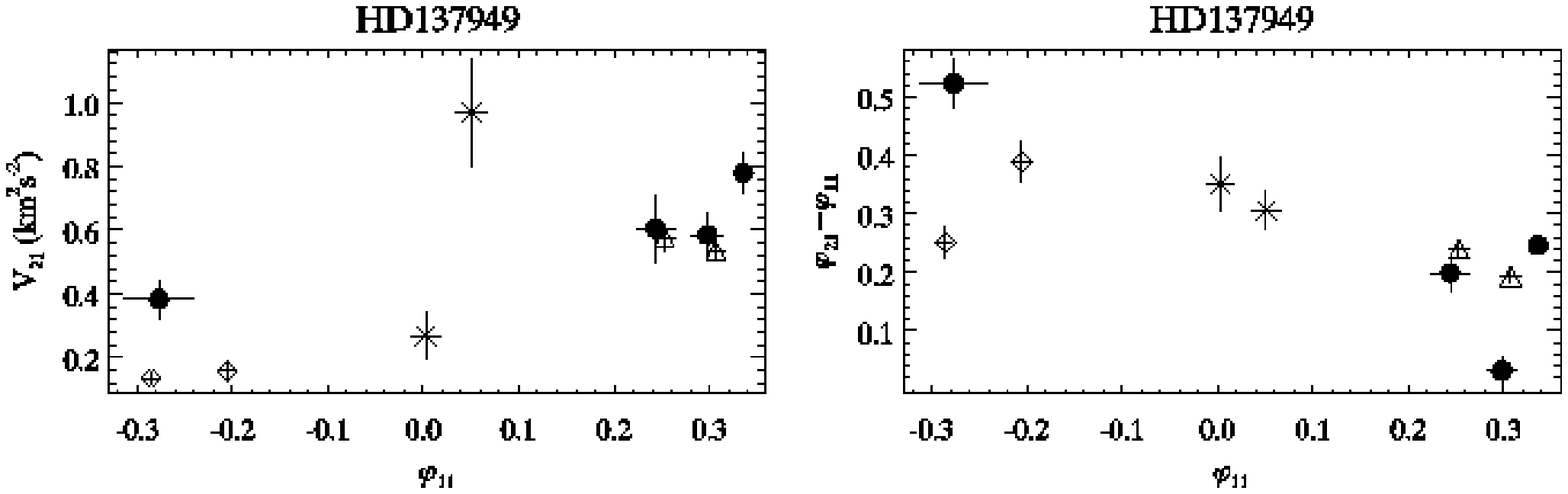}
\figps{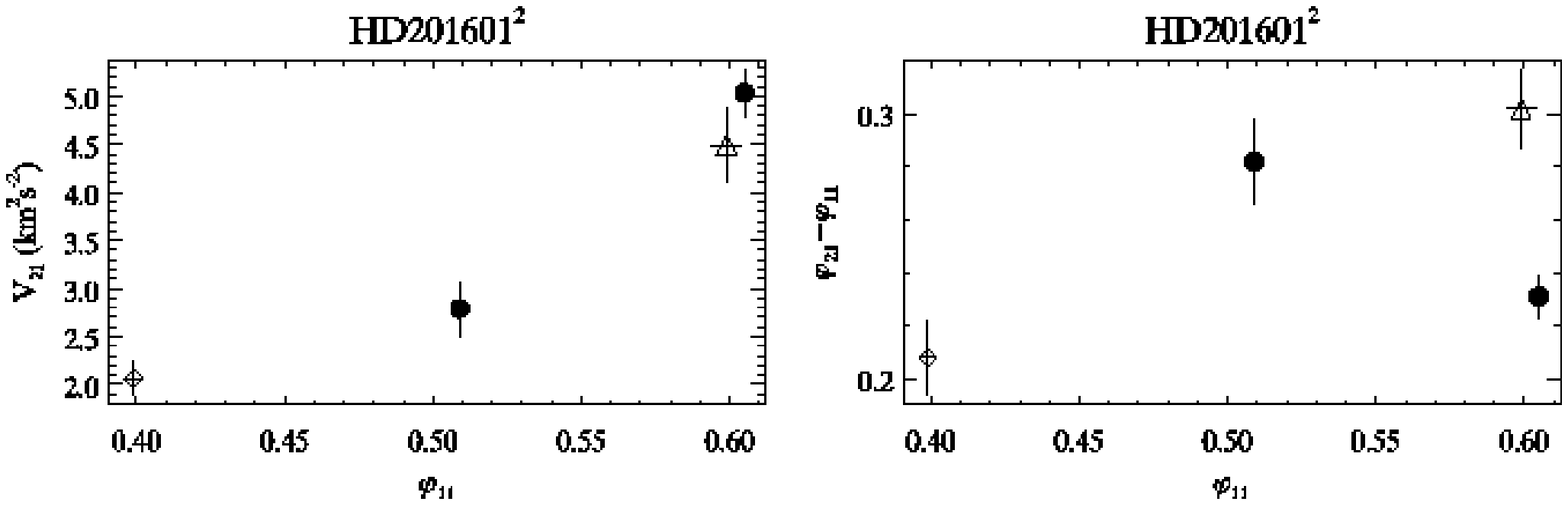}
\caption{Same as Fig.~\ref{fig8} for HD\,122970, HD\,128898, HD\,137949 and HD\,201601.}
\label{fig9}
\end{figure}

\begin{figure}
\fifps{7cm}{0}{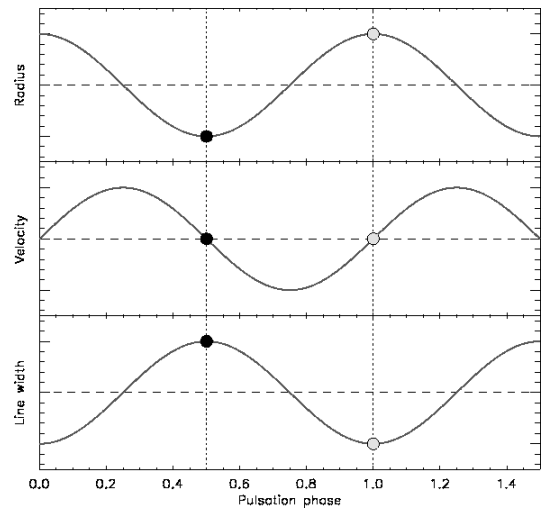}
\caption{Illustration of the phase relation between pulsational modulation of stellar radius, radial
velocity and line width discovered in our study. Symbols mark the phase point of maximum compression
(filled circle) and maximum expansion (open circle).} 
\label{fig10}
\end{figure}

Following arguments presented in Sect.~\ref{lines}, we use the phase $\varphi_{11}$ of  RV
fluctuations as a proxy of the relative formation height for REE lines. For pulsation waves propagating
outwards in the radial direction, a larger RV phase corresponds to a later time of maximum
velocity, which suggests a higher physical location of the line-forming region of the  REE
ion.

For all stars except HD\,137949 we discover a dramatic change of the second moment amplitude  $V_{21}$
with height. This trend is evident from inspection of Fig.~\ref{fig7} for HD\,19918 and HD\,101065.
The \ndt\ lines from deeper layers often exhibit large, or at least measurable harmonic variation of
\vb. As the pulsation wave propagates towards higher layers, the $V_{21}$ amplitude increases
steeply and the shape of \vb\ phase curve transforms from a low-amplitude, predominantly double wave to
a large-amplitude, mostly single wave. This tendency is  summarized for all stars in Figs.~\ref{fig8}
and \ref{fig9}. We find a persistent increase of the principal frequency component in the \vb\ variation
with increasing phase $\varphi_{11}$. Typically, \ndt\ lines have low $V_{21}$ amplitude; this
parameter increases in strong \nd\ lines, continues increasing for \pr\ and finally reaches maximum in
\tb\ (HD\,12932, HD\,24712, HD\,101065, HD\,122970, HD\,128898, HD\,134214). For a few stars (HD\,9289,
HD\,19918 and HD\,201601) where \tb\ lines do not show large-amplitude variability of \vb, the $V_{21}$
amplitude maximum occurs in \pr\ lines.

Comparing the inferred depth dependence of the second moment variation with the line-to-line changes
of LPV pattern, we discovered a remarkable correspondence between high-amplitude, single-wave
pulsational variability of \vb\ and the presence of a blue-to-red wave pattern in the LPV. Detailed 
intercomparison of the residual profile time series displayed in Figs.~\ref{fig3}--\ref{fig6} and
the moment fitting results reveals that all those lines, in particular \pr\ and \tb, showing
asymmetric LPV are invariably characterized by a large $V_{21}$ amplitude and high
$V_{21}/V_{11}^2$  ratio. The latter parameter reaches a value of 10--15 for \pr\ and \tb\ in
most stars. The depth variation of LPV pattern and $V_{21}$ amplitude is also consistent. The two stars,
HD\,101065 and HD\,122970, that develop asymmetric LPV late and show it prominently only in the \tb\
lines also exhibit less persistent single-wave second moment changes and have lower
$V_{21}/V_{11}^2$ ratios (2--7). Thus, an important conclusion emerging from our analysis of the first and
second moment variation with pulsation phase is that a large-amplitude, single-wave behaviour of
\vb\ and asymmetric LPV pattern are closely related and probably have the same physical origin.
We stress that this correlation between the two distinct characteristics of the roAp spectroscopic
variability is universal and is found in all stars where intrinsic harmonic variability is low
enough not to distort the moment technique diagnosis.
Consequently, any theory purporting to explain the asymmetric LPV pattern of sharp-lined roAp stars
must also reproduce the main features of the moment behaviour.

Further insight into pulsational line width perturbations is obtained by considering the phase
shift between the dominant single-wave variability of \va\ and \vb. In Figs.~\ref{fig8} and
\ref{fig9} we plot the phase delay, $\varphi_{21}-\varphi_{11}$, of the second moment with respect
to the phase $\varphi_{11}$ of the RV curve. The value of this phase lag is diverse for different groups of spectral lines,
and no consistent depth dependence is found. However, an interesting coherency is found for
the numerical value of $\varphi_{21}-\varphi_{11}$ in lines with dominant single-wave variation of the 
second moment. Using lines of \pr\ and \tb\ (for HD\,201601 the \nd\ lines are also considered) which
are sensitive to pulsations in the outermost stellar layers, we find that $\varphi_{21}-\varphi_{11}$
is always positive (see average values of the phase lag listed in Tables~\ref{tbl2}--\ref{tbl5}).
This means that the \vb\ maximum occurs later in time compared to the RV maximum. Furthermore,
the $\varphi_{21}-\varphi_{11}$ parameter spans a small range of values between 0.19 and 0.28. (The few
exceptions include HD\,9289, which has $\langle\varphi_{21}-\varphi_{11}\rangle=0.09$, and one
\equ\ data set with $\langle\varphi_{21}-\varphi_{11}\rangle=0.15$). For seven out of ten
stars studied, $\langle\varphi_{21}-\varphi_{11}\rangle$ is consistent within 1$\sigma$ confidence level
with exactly one-quarter period phase lag between RV and \vb. For two stars (HD\,24712 and
HD\,134214) the phase lag is consistent with the value of 0.25 at a 2--3$\sigma$ level.

\begin{figure}
\fifps{7cm}{0}{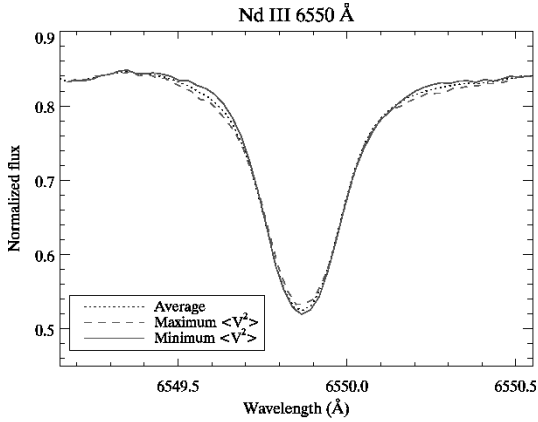}
\caption{Line width variation for the \nd\ 6550~\AA\ line in the spectrum of \equ. The average
profile (dotted curve) is compared with the time-resolved spectra corresponding to minimum 
(solid curve) and maximum (dashed curve) of the second moment.} 
\label{fig11}
\end{figure}

How does one interpret a systematic quarter-period delay between the variation of RV and line width? We
recall that, in the absence of significant non-linearities and non-adiabatic effects in stellar
pulsations, the velocity is shifted by 0.25 of the period with respect to radius changes. The negative extremum of \va\ corresponds to the middle of the expansion part of the radius
phase curve. At the maximum radius, RV returns to zero. Then the positive part of the RV curve describes
contraction phase, and radius reaches minimum when \va\ returns to zero after positive extremum. Adding
to this picture the \vb\ variability discussed above, we find that line width changes approximately in
antiphase (i.e., with a phase shift of 0.5) with respect to stellar radius variation. In other words,
the maximum contraction (minimum radius) phase point corresponds to maximum line width, whereas at the phase
of maximum expansion (maximum radius) lines become narrower. This fundamental relation between
pulsational modulation of line width on the one hand and RV and radius changes on the other hand
is sketched in Fig.~\ref{fig10}. 

The tendency of REE lines to become broader at the maximum contraction phase is confirmed by
investigation of the time-resolved spectra of \equ. As was already mentioned above, our 2001
CFHT observations caught this star in a remarkably high-amplitude pulsation state. The very high $S/N$
and high resolution of these data reveal pulsational profile variation synchronized
with radius changes. Fig.~\ref{fig11} compares the average profile of the \nd\ 6550~\AA\ line with the
time-resolved spectra corresponding to nearly zero RV, but different extrema of the \vb\ phase
curve. Time-resolved spectra for the minimum and maximum \vb\ exhibit different width,  yet the
equivalent width of the line does not change appreciably. This evidence confirms that our
measurements of the second moment of REE in roAp stars diagnose pulsational perturbation of line
width, not some other unrecognized line profile or line strength fluctuations.

\begin{table*}
\caption{Results of the least-squares fitting of the first and second moment variation of the
REE lines in HD\,9289, HD\,12932, HD\,19918, HD\,24712 and HD\,101065. The columns give line
identification, amplitude and phase ($V_{11}$, $\varphi_{11}$) for the \va\ variation at the 
main frequency and amplitude of the \va\ variability ($V_{12}$) at the first harmonic. The same 
quantities are listed for the second moment \vb. Amplitudes are measured in \kms\ for \va\
and in \kmq\ for \vb, phases
are given in units of the pulsation period. The HD number of the star is followed by the
phase difference $\varphi_{21}-\varphi_{11}$, estimated using the lines of \pr\ and \tb.
\label{tbl3}}
\begin{tabular}{lccccccc}
\hline 
~Ion   & $\lambda$ & \multicolumn{3}{c}{\va}                            &
                     \multicolumn{3}{c}{\vb}                            \\
       & (\AA)     & $V_{11}$ (\kms) & $\varphi_{11}$ & $V_{12}$ (\kms) &
                     $V_{21}$ (\kmq) & $\varphi_{21}$ & $V_{22}$ (\kmq) \\
\hline
\multicolumn{8}{c}{HD\,9289, $\langle\varphi_{21}-\varphi_{11}\rangle=0.09\pm0.04$} \\
\pr\  & 5284 & $0.549\pm0.024$ & $0.673\pm0.007$ & $0.048\pm0.024$ & $3.738\pm0.315$ & $0.742\pm0.013$ & $0.720\pm0.316$ \\
\nd\  & 5286 & $0.518\pm0.036$ & $0.524\pm0.011$ & $0.024\pm0.036$ & $3.087\pm0.348$ & $0.501\pm0.018$ & $0.143\pm0.350$ \\
\nd\  & 5294 & $0.603\pm0.031$ & $0.712\pm0.008$ & $0.068\pm0.031$ & $4.934\pm0.498$ & $0.786\pm0.016$ & $0.675\pm0.497$ \\
\pr\  & 5300 & $0.750\pm0.033$ & $0.689\pm0.007$ & $0.117\pm0.033$ & $5.980\pm0.492$ & $0.755\pm0.013$ & $0.696\pm0.492$ \\
\ndt\ & 5311 & $0.487\pm0.086$ & $0.584\pm0.028$ & $0.016\pm0.087$ & $3.921\pm0.863$ & $0.352\pm0.035$ & $1.482\pm0.862$ \\
\ndt\ & 5319 & $0.629\pm0.057$ & $0.538\pm0.014$ & $0.121\pm0.057$ & $2.890\pm0.581$ & $0.486\pm0.032$ & $1.013\pm0.583$ \\
\tb\  & 5505 & $0.533\pm0.029$ & $0.676\pm0.009$ & $0.057\pm0.029$ & $3.274\pm0.306$ & $0.744\pm0.015$ & $0.562\pm0.306$ \\
\nd\  & 5802 & $0.596\pm0.035$ & $0.565\pm0.009$ & $0.029\pm0.035$ & $3.811\pm0.474$ & $0.571\pm0.020$ & $1.101\pm0.475$ \\
\tb\  & 5847 & $0.451\pm0.082$ & $0.641\pm0.029$ & $0.109\pm0.082$ & $2.531\pm0.755$ & $0.781\pm0.048$ & $1.209\pm0.752$ \\
\nd\  & 5851 & $0.653\pm0.043$ & $0.559\pm0.011$ & $0.025\pm0.043$ & $4.133\pm0.652$ & $0.564\pm0.025$ & $1.364\pm0.651$ \\
\multicolumn{8}{c}{HD\,12932, $\langle\varphi_{21}-\varphi_{11}\rangle=0.23\pm0.02$} \\
\pr\  & 5284 & $1.052\pm0.039$ & $0.210\pm0.006$ & $0.015\pm0.039$ & $7.129\pm0.562$ & $0.413\pm0.012$ & $1.022\pm0.558$ \\
\nd\  & 5286 & $0.960\pm0.040$ & $0.033\pm0.007$ & $0.022\pm0.040$ & $1.592\pm0.208$ & $0.285\pm0.021$ & $0.491\pm0.206$ \\
\nd\  & 5294 & $0.843\pm0.030$ & $0.223\pm0.006$ & $0.050\pm0.030$ & $6.290\pm0.339$ & $0.466\pm0.009$ & $0.276\pm0.335$ \\
\pr\  & 5300 & $1.199\pm0.037$ & $0.184\pm0.005$ & $0.082\pm0.037$ & $8.512\pm0.428$ & $0.435\pm0.008$ & $0.772\pm0.426$ \\
\ndt\ & 5311 & $0.734\pm0.042$ & $0.975\pm0.009$ & $0.088\pm0.042$ & $0.830\pm0.181$ & $0.394\pm0.035$ & $0.258\pm0.181$ \\
\ndt\ & 5319 & $1.012\pm0.032$ & $0.005\pm0.005$ & $0.023\pm0.031$ & $1.388\pm0.162$ & $0.415\pm0.018$ & $0.280\pm0.161$ \\
\tb\  & 5505 & $1.292\pm0.041$ & $0.239\pm0.005$ & $0.064\pm0.041$ & $8.589\pm0.375$ & $0.485\pm0.007$ & $0.494\pm0.374$ \\
\nd\  & 5802 & $1.135\pm0.032$ & $0.088\pm0.004$ & $0.046\pm0.032$ & $3.119\pm0.221$ & $0.381\pm0.011$ & $0.665\pm0.219$ \\
\tb\  & 5847 & $1.359\pm0.087$ & $0.223\pm0.010$ & $0.077\pm0.086$ & $8.381\pm0.721$ & $0.440\pm0.014$ & $0.889\pm0.730$ \\
\nd\  & 5851 & $1.317\pm0.050$ & $0.101\pm0.006$ & $0.087\pm0.050$ & $4.040\pm0.576$ & $0.363\pm0.022$ & $1.101\pm0.574$ \\
\tb\  & 6323 & $1.066\pm0.074$ & $0.224\pm0.011$ & $0.087\pm0.075$ & $3.643\pm0.436$ & $0.479\pm0.019$ & $0.511\pm0.435$ \\
\multicolumn{8}{c}{HD\,19918, $\langle\varphi_{21}-\varphi_{11}\rangle=0.22\pm0.04$} \\
\pr\  & 5284 & $0.941\pm0.041$ & $0.056\pm0.007$ & $0.031\pm0.041$ & $8.673\pm0.352$ & $0.238\pm0.006$ & $0.839\pm0.352$ \\
\nd\  & 5286 & $0.951\pm0.062$ & $0.863\pm0.010$ & $0.095\pm0.062$ & $2.306\pm0.344$ & $0.054\pm0.024$ & $0.848\pm0.340$ \\
\nd\  & 5294 & $0.683\pm0.031$ & $0.070\pm0.007$ & $0.047\pm0.031$ & $7.613\pm0.374$ & $0.273\pm0.008$ & $0.353\pm0.374$ \\
\pr\  & 5300 & $1.184\pm0.054$ & $0.061\pm0.007$ & $0.051\pm0.053$ & $9.722\pm0.547$ & $0.285\pm0.009$ & $0.978\pm0.546$ \\
\ndt\ & 5311 & $1.087\pm0.074$ & $0.835\pm0.011$ & $0.129\pm0.073$ & $1.863\pm0.345$ & $0.097\pm0.030$ & $0.571\pm0.343$ \\
\ndt\ & 5319 & $1.225\pm0.044$ & $0.852\pm0.006$ & $0.140\pm0.044$ & $2.433\pm0.223$ & $0.188\pm0.014$ & $1.133\pm0.221$ \\
\nd\  & 5802 & $1.051\pm0.032$ & $0.927\pm0.005$ & $0.068\pm0.032$ & $3.632\pm0.219$ & $0.216\pm0.010$ & $1.117\pm0.219$ \\
\tb\  & 5847 & $0.887\pm0.076$ & $0.100\pm0.013$ & $0.091\pm0.075$ & $4.390\pm0.452$ & $0.354\pm0.017$ & $0.486\pm0.456$ \\
\nd\  & 5851 & $1.198\pm0.042$ & $0.928\pm0.006$ & $0.062\pm0.042$ & $4.022\pm0.249$ & $0.187\pm0.010$ & $0.968\pm0.248$ \\
\multicolumn{8}{c}{HD\,24712, $\langle\varphi_{21}-\varphi_{11}\rangle=0.19\pm0.02$} \\
\pr\  & 5284 & $0.265\pm0.015$ & $0.125\pm0.009$ & $0.001\pm0.015$ & $0.957\pm0.054$ & $0.337\pm0.009$ & $0.045\pm0.056$ \\
\nd\  & 5286 & $0.251\pm0.011$ & $0.833\pm0.007$ & $0.011\pm0.010$ & $0.476\pm0.045$ & $0.962\pm0.015$ & $0.058\pm0.044$ \\
\nd\  & 5294 & $0.176\pm0.009$ & $0.996\pm0.008$ & $0.007\pm0.009$ & $0.343\pm0.024$ & $0.209\pm0.011$ & $0.024\pm0.024$ \\
\pr\  & 5300 & $0.246\pm0.013$ & $0.114\pm0.009$ & $0.006\pm0.013$ & $0.913\pm0.056$ & $0.318\pm0.010$ & $0.039\pm0.056$ \\
\ndt\ & 5311 & $0.219\pm0.009$ & $0.821\pm0.007$ & $0.009\pm0.009$ & $0.275\pm0.032$ & $0.987\pm0.019$ & $0.019\pm0.033$ \\
\ndt\ & 5319 & $0.245\pm0.010$ & $0.851\pm0.007$ & $0.004\pm0.010$ & $0.280\pm0.026$ & $0.052\pm0.015$ & $0.069\pm0.027$ \\
\tb\  & 5505 & $0.385\pm0.026$ & $0.378\pm0.011$ & $0.009\pm0.025$ & $1.812\pm0.121$ & $0.575\pm0.011$ & $0.121\pm0.122$ \\
\tb\  & 5847 & $0.385\pm0.027$ & $0.348\pm0.011$ & $0.016\pm0.027$ & $1.820\pm0.189$ & $0.508\pm0.017$ & $0.216\pm0.190$ \\
\nd\  & 5851 & $0.243\pm0.011$ & $0.884\pm0.007$ & $0.008\pm0.011$ & $0.556\pm0.041$ & $0.990\pm0.012$ & $0.039\pm0.041$ \\
\multicolumn{8}{c}{HD\,101065, $\langle\varphi_{21}-\varphi_{11}\rangle=0.23\pm0.02$} \\
\pr\  & 5284 & $0.441\pm0.004$ & $0.943\pm0.001$ & $0.024\pm0.004$ & $0.258\pm0.017$ & $0.147\pm0.010$ & $0.204\pm0.017$ \\
\nd\  & 5286 & $0.278\pm0.005$ & $0.851\pm0.003$ & $0.015\pm0.005$ & $0.082\pm0.031$ & $0.888\pm0.059$ & $0.113\pm0.031$ \\
\nd\  & 5294 & $0.304\pm0.004$ & $0.897\pm0.002$ & $0.014\pm0.004$ & $0.277\pm0.023$ & $0.876\pm0.013$ & $0.202\pm0.023$ \\
\pr\  & 5300 & $0.460\pm0.005$ & $0.954\pm0.002$ & $0.031\pm0.005$ & $0.628\pm0.029$ & $0.185\pm0.007$ & $0.163\pm0.029$ \\
\ndt\ & 5311 & $0.215\pm0.003$ & $0.849\pm0.002$ & $0.009\pm0.003$ & $0.046\pm0.013$ & $0.619\pm0.047$ & $0.114\pm0.013$ \\
\ndt\ & 5319 & $0.189\pm0.002$ & $0.857\pm0.002$ & $0.008\pm0.002$ & $0.295\pm0.009$ & $0.356\pm0.005$ & $0.069\pm0.009$ \\
\nd\  & 5802 & $0.218\pm0.004$ & $0.855\pm0.003$ & $0.011\pm0.004$ & $0.227\pm0.014$ & $0.833\pm0.009$ & $0.098\pm0.014$ \\
\tb\  & 5847 & $0.704\pm0.013$ & $0.111\pm0.003$ & $0.028\pm0.013$ & $3.588\pm0.128$ & $0.360\pm0.006$ & $0.089\pm0.128$ \\
\nd\  & 5851 & $0.239\pm0.005$ & $0.847\pm0.003$ & $0.013\pm0.005$ & $0.281\pm0.024$ & $0.822\pm0.013$ & $0.166\pm0.023$ \\
\tb\  & 6323 & $0.646\pm0.015$ & $0.107\pm0.004$ & $0.037\pm0.015$ & $2.856\pm0.124$ & $0.337\pm0.007$ & $0.087\pm0.123$ \\
\hline                               
\end{tabular}                        
\end{table*}

\begin{table*}
\caption{Results of the least-squares fitting of the first and second moment variation of the
REE lines in HD\,122970, HD\,128898, HD\,134214 and HD\,137949. See caption of Table~\ref{tbl3}
for the explanation of this table's content. \label{tbl4}}
\begin{tabular}{lccccccc}
\hline 
~Ion   & $\lambda$ & \multicolumn{3}{c}{\va}                            &
                     \multicolumn{3}{c}{\vb}                            \\
       & (\AA)     & $V_{11}$ (\kms) & $\varphi_{11}$ & $V_{12}$ (\kms) &
                     $V_{21}$ (\kmq) & $\varphi_{21}$ & $V_{22}$ (\kmq) \\
\hline
\multicolumn{8}{c}{HD\,122970, $\langle\varphi_{21}-\varphi_{11}\rangle=0.22\pm0.04$} \\
\pr\  & 5284 & $0.614\pm0.016$ & $0.891\pm0.004$ & $0.032\pm0.016$ & $0.930\pm0.068$ & $0.926\pm0.012$ & $0.465\pm0.068$ \\
\nd\  & 5286 & $0.535\pm0.027$ & $0.844\pm0.008$ & $0.021\pm0.027$ & $0.309\pm0.120$ & $0.729\pm0.063$ & $0.227\pm0.121$ \\
\nd\  & 5294 & $0.434\pm0.011$ & $0.853\pm0.004$ & $0.025\pm0.011$ & $0.270\pm0.040$ & $0.875\pm0.024$ & $0.118\pm0.040$ \\
\pr\  & 5300 & $0.586\pm0.014$ & $0.886\pm0.004$ & $0.020\pm0.014$ & $0.283\pm0.058$ & $0.086\pm0.033$ & $0.245\pm0.058$ \\
\ndt\ & 5311 & $0.490\pm0.023$ & $0.842\pm0.007$ & $0.028\pm0.023$ & $0.103\pm0.101$ & $0.828\pm0.154$ & $0.154\pm0.101$ \\
\ndt\ & 5319 & $0.430\pm0.012$ & $0.843\pm0.005$ & $0.006\pm0.012$ & $0.074\pm0.041$ & $0.206\pm0.089$ & $0.163\pm0.041$ \\
\tb\  & 5505 & $1.107\pm0.026$ & $0.953\pm0.004$ & $0.136\pm0.026$ & $1.504\pm0.146$ & $0.213\pm0.015$ & $0.838\pm0.145$ \\
\nd\  & 5802 & $0.511\pm0.020$ & $0.846\pm0.006$ & $0.019\pm0.020$ & $0.126\pm0.085$ & $0.953\pm0.108$ & $0.143\pm0.085$ \\
\tb\  & 5847 & $1.095\pm0.039$ & $0.930\pm0.006$ & $0.113\pm0.038$ & $1.383\pm0.210$ & $0.113\pm0.024$ & $1.185\pm0.211$ \\
\nd\  & 5851 & $0.515\pm0.019$ & $0.853\pm0.006$ & $0.030\pm0.019$ & $0.553\pm0.096$ & $0.948\pm0.028$ & $0.422\pm0.096$ \\
\tb\  & 6323 & $1.197\pm0.053$ & $0.946\pm0.007$ & $0.169\pm0.052$ & $1.239\pm0.320$ & $0.176\pm0.041$ & $1.096\pm0.315$ \\
\multicolumn{8}{c}{HD\,128898, $\langle\varphi_{21}-\varphi_{11}\rangle=0.28\pm0.04$} \\
\pr\  & 5284 & $0.509\pm0.011$ & $0.670\pm0.003$ & $0.017\pm0.011$ & $2.715\pm0.084$ & $0.887\pm0.005$ & $0.046\pm0.084$ \\
\nd\  & 5286 & $0.708\pm0.018$ & $0.359\pm0.004$ & $0.075\pm0.018$ & $2.248\pm0.132$ & $0.687\pm0.009$ & $0.346\pm0.132$ \\
\nd\  & 5294 & $0.287\pm0.009$ & $0.604\pm0.005$ & $0.025\pm0.009$ & $1.740\pm0.080$ & $0.911\pm0.007$ & $0.002\pm0.080$ \\
\pr\  & 5300 & $0.513\pm0.012$ & $0.660\pm0.004$ & $0.020\pm0.012$ & $2.916\pm0.100$ & $0.937\pm0.005$ & $0.133\pm0.099$ \\
\ndt\ & 5319 & $0.815\pm0.018$ & $0.333\pm0.003$ & $0.067\pm0.018$ & $2.553\pm0.132$ & $0.778\pm0.008$ & $0.289\pm0.131$ \\
\tb\  & 5505 & $0.777\pm0.017$ & $0.840\pm0.003$ & $0.059\pm0.017$ & $5.088\pm0.153$ & $0.137\pm0.005$ & $0.014\pm0.154$ \\
\nd\  & 5802 & $0.512\pm0.013$ & $0.432\pm0.004$ & $0.036\pm0.013$ & $1.930\pm0.110$ & $0.798\pm0.009$ & $0.236\pm0.110$ \\
\tb\  & 5847 & $0.854\pm0.033$ & $0.834\pm0.006$ & $0.076\pm0.033$ & $5.376\pm0.257$ & $0.123\pm0.008$ & $0.295\pm0.257$ \\
\nd\  & 5851 & $0.563\pm0.014$ & $0.441\pm0.004$ & $0.046\pm0.014$ & $2.088\pm0.105$ & $0.769\pm0.008$ & $0.431\pm0.105$ \\
\tb\  & 6323 & $0.746\pm0.040$ & $0.802\pm0.008$ & $0.055\pm0.040$ & $3.781\pm0.271$ & $0.108\pm0.011$ & $0.612\pm0.270$ \\
\multicolumn{8}{c}{HD\,134214, $\langle\varphi_{21}-\varphi_{11}\rangle=0.20\pm0.03$} \\
\pr\  & 5284 & $0.387\pm0.016$ & $0.226\pm0.007$ & $0.010\pm0.016$ & $2.241\pm0.102$ & $0.376\pm0.007$ & $0.063\pm0.103$ \\
\nd\  & 5286 & $0.553\pm0.021$ & $0.921\pm0.006$ & $0.014\pm0.021$ & $1.233\pm0.127$ & $0.099\pm0.016$ & $0.100\pm0.128$ \\
\nd\  & 5294 & $0.207\pm0.010$ & $0.095\pm0.007$ & $0.010\pm0.010$ & $0.728\pm0.052$ & $0.374\pm0.011$ & $0.052\pm0.052$ \\
\pr\  & 5300 & $0.362\pm0.015$ & $0.205\pm0.006$ & $0.013\pm0.015$ & $2.266\pm0.104$ & $0.425\pm0.007$ & $0.122\pm0.104$ \\
\ndt\ & 5311 & $0.508\pm0.019$ & $0.894\pm0.006$ & $0.001\pm0.020$ & $0.482\pm0.073$ & $0.120\pm0.024$ & $0.242\pm0.072$ \\
\ndt\ & 5319 & $0.522\pm0.018$ & $0.928\pm0.006$ & $0.010\pm0.018$ & $0.843\pm0.053$ & $0.214\pm0.010$ & $0.155\pm0.053$ \\
\tb\  & 5505 & $0.508\pm0.027$ & $0.410\pm0.009$ & $0.026\pm0.027$ & $2.713\pm0.159$ & $0.614\pm0.009$ & $0.143\pm0.158$ \\
\nd\  & 5802 & $0.444\pm0.018$ & $0.963\pm0.006$ & $0.004\pm0.018$ & $1.137\pm0.085$ & $0.155\pm0.012$ & $0.137\pm0.085$ \\
\tb\  & 5847 & $0.549\pm0.043$ & $0.395\pm0.013$ & $0.017\pm0.043$ & $4.243\pm0.375$ & $0.613\pm0.014$ & $0.513\pm0.377$ \\
\nd\  & 5851 & $0.490\pm0.019$ & $0.964\pm0.006$ & $0.019\pm0.019$ & $1.308\pm0.098$ & $0.179\pm0.012$ & $0.116\pm0.099$ \\
\tb\  & 6323 & $0.673\pm0.060$ & $0.413\pm0.014$ & $0.045\pm0.059$ & $3.683\pm0.546$ & $0.606\pm0.023$ & $0.527\pm0.549$ \\
\multicolumn{8}{c}{HD\,137949, $\langle\varphi_{21}-\varphi_{11}\rangle=0.27\pm0.07$} \\
\pr\  & 5284 & $0.056\pm0.004$ & $0.307\pm0.010$ & $0.028\pm0.004$ & $0.535\pm0.027$ & $0.500\pm0.008$ & $0.242\pm0.027$ \\
\nd\  & 5286 & $0.023\pm0.005$ & $0.723\pm0.036$ & $0.059\pm0.005$ & $0.383\pm0.060$ & $0.246\pm0.025$ & $0.588\pm0.059$ \\
\nd\  & 5294 & $0.128\pm0.006$ & $0.336\pm0.007$ & $0.011\pm0.006$ & $0.779\pm0.064$ & $0.582\pm0.013$ & $0.246\pm0.064$ \\
\pr\  & 5300 & $0.069\pm0.005$ & $0.253\pm0.011$ & $0.035\pm0.005$ & $0.576\pm0.044$ & $0.492\pm0.012$ & $0.293\pm0.044$ \\
\ndt\ & 5311 & $0.096\pm0.005$ & $0.715\pm0.007$ & $0.047\pm0.005$ & $0.135\pm0.020$ & $0.966\pm0.024$ & $0.499\pm0.020$ \\
\ndt\ & 5319 & $0.114\pm0.005$ & $0.795\pm0.008$ & $0.066\pm0.005$ & $0.157\pm0.034$ & $0.184\pm0.035$ & $0.642\pm0.034$ \\
\nd\  & 5802 & $0.064\pm0.006$ & $0.298\pm0.016$ & $0.042\pm0.006$ & $0.585\pm0.065$ & $0.330\pm0.018$ & $0.228\pm0.065$ \\
\tb\  & 5847 & $0.096\pm0.009$ & $0.003\pm0.015$ & $0.062\pm0.009$ & $0.268\pm0.072$ & $0.355\pm0.043$ & $0.587\pm0.072$ \\
\nd\  & 5851 & $0.064\pm0.008$ & $0.244\pm0.020$ & $0.070\pm0.008$ & $0.604\pm0.103$ & $0.443\pm0.027$ & $0.532\pm0.103$ \\
\tb\  & 6323 & $0.175\pm0.017$ & $0.051\pm0.016$ & $0.099\pm0.017$ & $0.971\pm0.169$ & $0.357\pm0.028$ & $0.929\pm0.169$ \\
\hline                               
\end{tabular}                        
\end{table*}

\begin{table*}
\caption{Results of the least-squares fitting of the first and second moment variation of the
REE lines in HD\,201601. See caption of Table~\ref{tbl3} for the explanation of this table's content. 
The phase difference $\varphi_{21}-\varphi_{11}$ is obtained using the lines of \pr, \nd\ and \tb.
\label{tbl5}}
\begin{tabular}{lccccccc}
\hline 
~Ion   & $\lambda$ & \multicolumn{3}{c}{\va}                            &
                     \multicolumn{3}{c}{\vb}                            \\
       & (\AA)     & $V_{11}$ (\kms) & $\varphi_{11}$ & $V_{12}$ (\kms) &
                     $V_{21}$ (\kmq) & $\varphi_{21}$ & $V_{22}$ (\kmq) \\
\hline
\multicolumn{8}{c}{HD\,201601$^1$, $\langle\varphi_{21}-\varphi_{11}\rangle=0.21$} \\
\nd\  & 6145 & $0.458\pm0.016$ & $0.178\pm0.005$ & $0.056\pm0.016$ & $2.989\pm0.186$ & $0.390\pm0.010$ & $0.574\pm0.194$ \\
\pr\  & 6160 & $0.642\pm0.018$ & $0.179\pm0.004$ & $0.067\pm0.019$ & $3.192\pm0.183$ & $0.385\pm0.009$ & $0.302\pm0.188$ \\
\multicolumn{8}{c}{HD\,201601$^2$, $\langle\varphi_{21}-\varphi_{11}\rangle=0.26\pm0.05$} \\
\nd\  & 6550 & $0.737\pm0.011$ & $0.605\pm0.002$ & $0.060\pm0.011$ & $5.038\pm0.248$ & $0.836\pm0.008$ & $0.066\pm0.246$ \\
\ndt\ & 6650 & $1.131\pm0.021$ & $0.399\pm0.003$ & $0.130\pm0.020$ & $2.063\pm0.178$ & $0.607\pm0.014$ & $0.969\pm0.177$ \\
\nd\  & 6691 & $1.078\pm0.020$ & $0.509\pm0.003$ & $0.097\pm0.020$ & $2.792\pm0.280$ & $0.791\pm0.016$ & $1.193\pm0.276$ \\
\pr\  & 6706 & $0.841\pm0.027$ & $0.599\pm0.005$ & $0.070\pm0.027$ & $4.495\pm0.392$ & $0.901\pm0.014$ & $1.081\pm0.393$ \\
\multicolumn{8}{c}{HD\,201601$^3$, $\langle\varphi_{21}-\varphi_{11}\rangle=0.26\pm0.06$} \\
\pr\  & 5284 & $0.158\pm0.006$ & $0.345\pm0.006$ & $0.007\pm0.006$ & $0.405\pm0.064$ & $0.581\pm0.025$ & $0.075\pm0.064$ \\
\nd\  & 5286 & $0.319\pm0.013$ & $0.185\pm0.007$ & $0.011\pm0.013$ & $0.714\pm0.144$ & $0.431\pm0.032$ & $0.079\pm0.143$ \\
\nd\  & 5294 & $0.104\pm0.007$ & $0.372\pm0.011$ & $0.002\pm0.007$ & $0.791\pm0.067$ & $0.564\pm0.014$ & $0.101\pm0.068$ \\
\pr\  & 5300 & $0.234\pm0.010$ & $0.346\pm0.006$ & $0.025\pm0.010$ & $1.094\pm0.114$ & $0.577\pm0.017$ & $0.152\pm0.116$ \\
\ndt\ & 5311 & $0.277\pm0.013$ & $0.139\pm0.008$ & $0.032\pm0.013$ & $0.253\pm0.056$ & $0.473\pm0.036$ & $0.085\pm0.056$ \\
\ndt\ & 5319 & $0.310\pm0.010$ & $0.178\pm0.005$ & $0.018\pm0.010$ & $0.303\pm0.042$ & $0.465\pm0.022$ & $0.095\pm0.042$ \\
\tb\  & 5847 & $0.222\pm0.038$ & $0.499\pm0.028$ & $0.071\pm0.038$ & $1.193\pm0.349$ & $0.851\pm0.047$ & $0.268\pm0.353$ \\
\nd\  & 5851 & $0.218\pm0.016$ & $0.375\pm0.012$ & $0.028\pm0.016$ & $0.667\pm0.130$ & $0.661\pm0.031$ & $0.290\pm0.129$ \\
\multicolumn{8}{c}{HD\,201601$^4$, $\langle\varphi_{21}-\varphi_{11}\rangle=0.15$} \\
\tb\  & 5847 & $0.178\pm0.046$ & $0.823\pm0.041$ & $0.034\pm0.046$ & $1.318\pm0.489$ & $0.997\pm0.059$ & $0.440\pm0.481$ \\
\nd\  & 5851 & $0.199\pm0.016$ & $0.705\pm0.013$ & $0.012\pm0.016$ & $0.515\pm0.144$ & $0.838\pm0.044$ & $0.250\pm0.144$ \\
\multicolumn{8}{c}{HD\,201601$^5$, $\langle\varphi_{21}-\varphi_{11}\rangle=0.25$} \\
\nd\  & 6550 & $0.118\pm0.013$ & $0.906\pm0.017$ & $0.002\pm0.013$ & $0.579\pm0.167$ & $0.182\pm0.046$ & $0.231\pm0.165$ \\
\ndt\ & 6650 & $0.149\pm0.028$ & $0.715\pm0.031$ & $0.015\pm0.029$ & $0.214\pm0.166$ & $0.933\pm0.123$ & $0.173\pm0.166$ \\
\multicolumn{8}{c}{HD\,201601$^6$, $\langle\varphi_{21}-\varphi_{11}\rangle=0.23$} \\
\nd\  & 6145 & $0.155\pm0.008$ & $0.521\pm0.008$ & $0.013\pm0.008$ & $0.843\pm0.148$ & $0.751\pm0.028$ & $0.102\pm0.147$ \\
\pr\  & 6160 & $0.175\pm0.011$ & $0.546\pm0.010$ & $0.025\pm0.011$ & $0.910\pm0.082$ & $0.783\pm0.014$ & $0.188\pm0.081$ \\
\hline                               
\end{tabular}                        
\end{table*}

\subsection{Short-term variation of moment amplitudes}

\citet{KR01b} discovered that the amplitude of the pulsational RV variability of the \nd\ 6145~\AA\
line in \cir\ (HD\,128898) changes significantly within a time period of about one hour. This
modulation of RV oscillations occurs on a time scale much shorter than the 4.471~d rotation period of
\cir\ and cannot be ascribed to beating between known close frequencies because photometric studies
\citep{KSM94} show that this roAp star is essentially a monoperiodic pulsator. In a recent
investigation of the pulsational RV variability of \pr\ lines, \citet*{KEM06} confirmed the presence of 
short-term changes of RV amplitude in \cir\ and demonstrated the existence of similar behaviour in a few
other roAp stars. Among various possible explanations for the new spectroscopic frequencies, observed
in \pr\ but absent in broad-band photometry, \citet*{KEM06} favoured the idea that at the formation
heights of \pr\ lines the pulsational RV amplitude experiences stochastic variation due to growth and
decay of the principal pulsation mode.

\begin{figure}
\figps{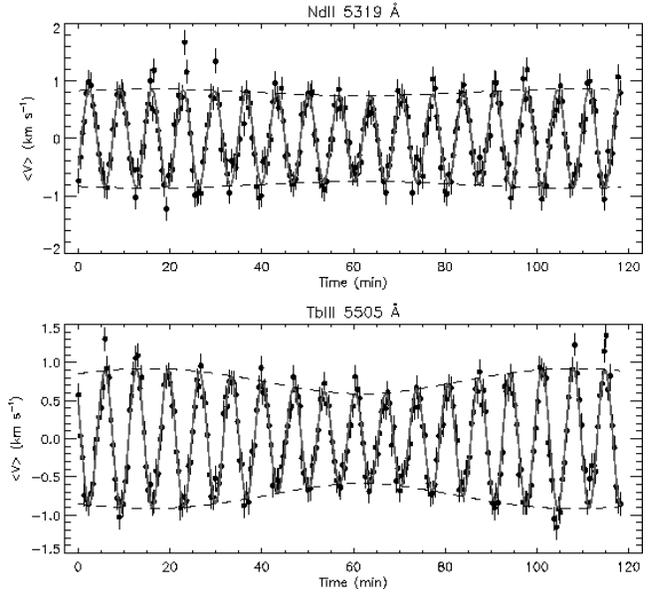}
\caption{Modulation of RV amplitude for the \ndt\ 5319~\AA\ (upper panel) and \tb\ 5505~\AA\
(lower panel) lines in HD\,128898. Symbols show velocity measurements, solid line corresponds to the
best-fitting superposition of the variability with $P_1=6.80$~min and $P_2=7.34$~min. The dashed curve
illustrates temporal variation of the principal frequency amplitude.} 
\label{fig12}
\end{figure}

We have performed a new analysis of the short-term changes of pulsational variability using
time series of both the first and the second moments measured in the UVES time-resolved spectra of
\cir. We explored the behaviour of several REE ions (\pr, \ndt, \nd, \tb) and considered the results in the
light of the newly discovered depth dependence of the pulsational LPV pattern and line width
oscillations. Analysis of the amplitude variability for the REE line profile moments was based on the
least-squares fitting introduced in Sect.~\ref{sect_moments}, except that here we adjust amplitude and
phase for the principal frequency ($P_1=6.80$~min) and a secondary periodicity $P_2=7.34$~min
\citep*{KEM06}, responsible for the observed short-term modulation. The strength of the amplitude
modulation is then characterized by the respective amplitude ratio $V_2/V_1$, estimated separately
for the first and second moments. 

\begin{table}
\caption{Amplitude modulation for the oscillations in the first and second moments of REE lines in \cir. 
The columns give line identification and the amplitude ratio $V_2/V_1$, which characterizes the strength
of the secondary periodicity ($P_2=7.34$~min) with respect to the amplitude of the principal 
frequency ($P_1=6.80$~min). \label{tbl6}}
\centering
\begin{tabular}{lccc}
\hline 
~Ion   & $\lambda$ & \multicolumn{2}{c}{$V_2/V_1$}  \\
       & (\AA)     & \va\          &           \vb\ \\
\hline
\pr\   & 5284 & $0.192\pm0.018$ & $0.331\pm0.026$ \\
\nd\   & 5286 & $0.154\pm0.025$ & $0.392\pm0.062$ \\
\nd\   & 5294 & $0.255\pm0.031$ & $0.360\pm0.046$ \\
\pr\   & 5300 & $0.165\pm0.022$ & $0.369\pm0.029$ \\
\ndt\  & 5319 & $0.073\pm0.022$ & $0.224\pm0.053$ \\
\tb\   & 5505 & $0.220\pm0.018$ & $0.371\pm0.023$ \\
\nd\   & 5802 & $0.141\pm0.024$ & $0.423\pm0.059$ \\
\tb\   & 5847 & $0.188\pm0.038$ & $0.334\pm0.050$ \\
\nd\   & 5851 & $0.198\pm0.024$ & $0.370\pm0.054$ \\
\hline                               
\end{tabular}                        
\end{table}

Numerical results of the least-squares analysis of the moment time series is summarized in
Table~\ref{tbl6} for nine REE lines. Clearly, we find different amounts of RV amplitude modulation
for different REE ions. The secondary frequency is at the detection threshold for the \ndt\ 5319~\AA\
line. The $V_2/V_1$ amplitude ratio increases by a factor of 2--3 towards the higher atmospheric
layers where doubly ionized REE lines are formed. This is further illustrated in Fig.~\ref{fig12},
where we show the observed and best-fitting RV curve for the \ndt\ 5319~\AA\ and \tb\ 5505~\AA\ lines.
One can clearly see that pulsations in the latter line change appreciably with time, whereas much weaker
modulation is observed for the former line. This figure and Table~\ref{tbl6} provides the
first definite demonstration of the height variation of the short-term amplitude modulation effect in
\cir. Moreover, time-series analysis of the second moment reveals a much stronger (by a factor
1.5--3) amplitude modulation compared to the RV results. This is unexpected because the relative
amplitudes of the two frequencies should be the same for both \va\ and \vb\ \citep{MAP94}. 

Thus, it is tempting to speculate that, whatever effect gives rise to the observed RV amplitude variability
in the upper atmosphere of \cir, it acts more strongly on the line width variation than on the first
moment. Considering these results in the light of our discovery of the steep growth of the line width
oscillations with height, we suggest that the observed short-term modulation is actually a property
of the physical mechanism responsible for the line width and asymmetric LPV behaviour, while the
short-term amplitude changes visible in RV represent a secondary consequence of the action of this
hypothetical mechanism. Further investigation of several roAp stars is required to verify this
hypothesis. In particular, a full night of uninterrupted time-resolved monitoring of \cir\ would be
extremely useful to discriminate between a truly stochastic and a quasiperiodic nature of the variation
of oscillation amplitudes in the upper atmospheric layers. 

\section{Interpretation}
\label{newopm}

Our survey of the pulsational line profile variability in sharp-lined roAp stars has revealed
a new type of LPV behaviour and has demonstrated its significant depth dependence. In summary, our
observational findings are:
\begin{enumerate}
\item The asymmetric, blue-to-red wave pattern is common in time-resolved residual spectra of roAp stars.
This behaviour is more pronounced for REE lines formed in the higher atmospheric layers,
whereas lines originating at the bottom of the REE-enriched cloud often display S-shaped,  
symmetric, blue-to-red-to-blue LPV pattern.
\item The doubly ionized REE lines from the uppermost part of the stellar atmosphere show strong single-wave
pulsational variability of line width. These oscillations 
typically occur in antiphase with the pulsational changes of stellar radius. They also 
correlate with the presence of asymmetric LPV and exhibit the same height dependence.
\item The efficiency of the mechanism responsible for the asymmetric LPV pattern and single-wave
line width variation in HD\,128898 changes on relatively short time scales.
\end{enumerate}

These results, in particular the new type of pulsational LPV described by (i) and (ii), challenge
our understanding of pulsations in roAp stars and call for theoretical explanation. In this
section we present a series of line profile calculations for oblique non-radial
pulsations and compare observations with theoretical expectations. We show that the standard
models including only spherical harmonic perturbations of velocity fail to reproduce the
observed line profile and moment variability. This led us to introduce a new oblique pulsator model,
which accounts for additional line width oscillations. Using this new modelling framework, we
are able to explain all the main characteristics of the pulsational LPV in sharp-lined roAp stars.

\subsection{Spectrum synthesis calculations for oblique pulsators}

Periodic variation of velocity in non-radial pulsators is described with the spherical harmonics
$Y^m_\ell(\theta,\phi)$ and their derivatives:
\begin{eqnarray}
 V_r      & = & V_p Y^m_\ell (\theta,\phi) {\rm e}^{{\rm i}\omega t}, \nonumber \\
 V_\theta & = & K V_p \dfrac{\partial}{\partial\theta} 
                  Y^m_\ell (\theta,\phi) {\rm e}^{{\rm i}\omega t},  \label{vel}\\
 V_\phi   & = & K V_p \dfrac{1}{\sin{\theta}}\dfrac{\partial}{\partial\phi} 
                  Y^m_\ell (\theta,\phi) {\rm e}^{{\rm i}\omega t},\nonumber
\end{eqnarray}
where $\omega$ is the angular pulsation frequency, $V_p$ is the pulsation amplitude and $K$ is
the ratio of the vertical to horizontal amplitudes. Non-radial oscillations in roAp stars are
oblique. Therefore, spherical angular coordinates $\theta, \phi$ refer to a reference system
whose orientation is defined by the angles $\beta$ and $\chi$ with respect to the stellar axis
of rotation. In the classical oblique pulsator model, the pulsation axis is assumed to be aligned
with the axis of a quasi-dipolar magnetic field. In this case, $\theta$ and $\phi$ are spherical
angular coordinates in the magnetic reference frame.

Eq.~(\ref{vel}) can easily be generalized \citep[see][]{K04a} for the situation when pulsation
modes are distorted by stellar rotation and/or a strong magnetic field and are described by
superposition of several spherical harmonic functions.

The horizontal to vertical amplitude ratio is usually approximated as
\beq
K = \dfrac{GM}{\omega^2 R^3}
\eeq
\citep[e.g.,][]{ST87}. This relation predicts that vertical oscillations should be dominant
(i.e., $K\ll 1$) for high-overtone $p$-modes in roAp stars. Consequently, horizontal
displacement was often neglected in previous studies of roAp pulsations \citep{BD02}. However, the distorted
magnetoacoustic pulsations modes in the theory of \citet{SG04} show significant 
horizontal amplitudes ($K\sim 1$) in the lower part of stellar atmosphere. 

For the purpose of computing theoretical LPV, the pulsation velocity field described by
Eq.~(\ref{vel})  is transformed to the Descartes coordinate system of the observer. The line of
sight velocity component, $V^{\rm o}$, which enters spectrum synthesis, combines the contribution of
pulsations and the Doppler shift due to solid body stellar rotation with the projected rotational
velocity \vs. For the sake of brevity here we omit details of coordinate transformations,
definition of spherical harmonics and calculation of their derivatives. Interested readers can
find this information in \citet{K04a,K05}.

Even neglecting possible rotational and magnetic distortion of the non-radial pulsations, the model
formulated above has a large number of free parameters. One has to specify the amplitude and
structure of the pulsation mode, $V_p$, $K$, $\ell$, $m$, as well as the orientation of the pulsation axis
$\beta$, $\chi$, and the stellar \vs\ and inclination angle $i$. In the limit of negligible
\vs\ applicable to many roAp stars, the angles $\beta$, $\chi$ and $i$ can be replaced by 
a single parameter $\alpha$ --  the angle between the line of sight and pulsation axis. 

In order to be able to explore properties of LPV for many different pulsation geometries in a
reasonable amount of time, we approximate the local line profile with a Gaussian function,
characterized by a constant central depth $D$ and full width at half maximum $W$.
Then the disk-integrated line profile $S$ is obtained by the weighted summation over all visible
surface zones
\beq
S = 1 - D \sum_{j=1}^N T_j {\rm e}^{{\displaystyle -4\ln{2}}\dfrac{\left(v-V^{\rm o}_j\right)^2}{W^2}},
\label{disk}
\eeq
where the weight function 
\beq
T_j = \dfrac{3 (1-u+u\mu_j)\mu_j A_j}{\pi (3-u)}
\eeq
accounts for unequal projected areas of surface elements and
includes a linear limb-darkening law with coefficient $u$, and $\mu$ denotes the 
cosine of the angle between the line of sight and the surface normal. In all simulations described
below we use $D=0.5$, $W=10$~\kms, $u=0.5$. $V_p$ and $K$ are chosen in such a way that the
resultant RV amplitude is compatible with the one typically found in roAp stars ($\sim 1$~\kms). The stellar
surface is sampled on a 3909-element grid with unequal number of surface zones at different
latitudes \citep[see][]{PK02}. Calculations are performed for a set of 20 equally-spaced pulsation
phases. Eq.~(\ref{moments}) is used to infer variation of the first and second moments 
from theoretical line profile time series.

Our LPV synthesis model is very similar to the one usually used in a broader context of spectral
variability studies of non-radial pulsations in slowly rotating stars \citep{AW93,STA97}. Using
this standard description of oscillations and accounting for the additional complication caused by
oblique mode geometry, we carried out an extensive set of computations aiming to find a
combination of input parameters that produces LPV compatible with observations. Pulsational LPV
and moment variability was investigated for
\begin{enumerate}
\renewcommand{\theenumi}{(\arabic{enumi})}
\item all possible modes with $\ell\le 3$;
\item projected rotational velocities in the range 0--10~\kms;
\item modes with significant ($K\sim 1$) and dominant ($K\gg 1$) horizontal fluctuations;
\item distorted dipolar modes given by superposition of $\ell=1$, $m=-1, 0, 1$ spherical
harmonics \citep{BD02} or by superposition of $\ell=1$ and $\ell=3$ axisymmetric harmonics
\citep{SG04};
\item modes containing two pulsation components with different $K$ and a phase shift relative 
to each other -- this model is meant to simulate superposition of magnetic and acoustic
running wave components possibly existing in the upper atmospheres of roAp stars \citep{C06}.
\end{enumerate}

\begin{figure*}
\fifps{8.5cm}{0}{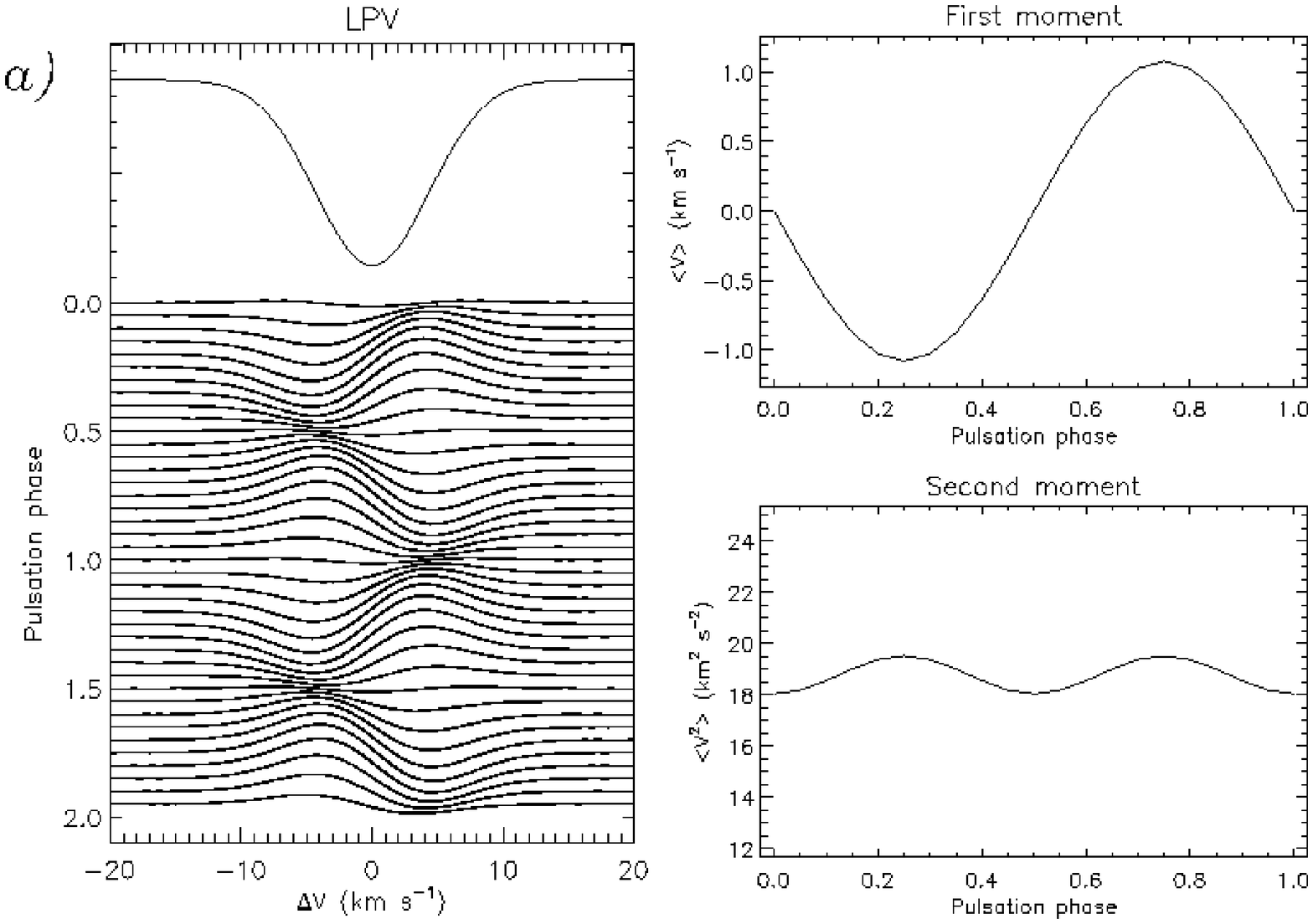}\hspace{0.2cm}
\fifps{8.5cm}{0}{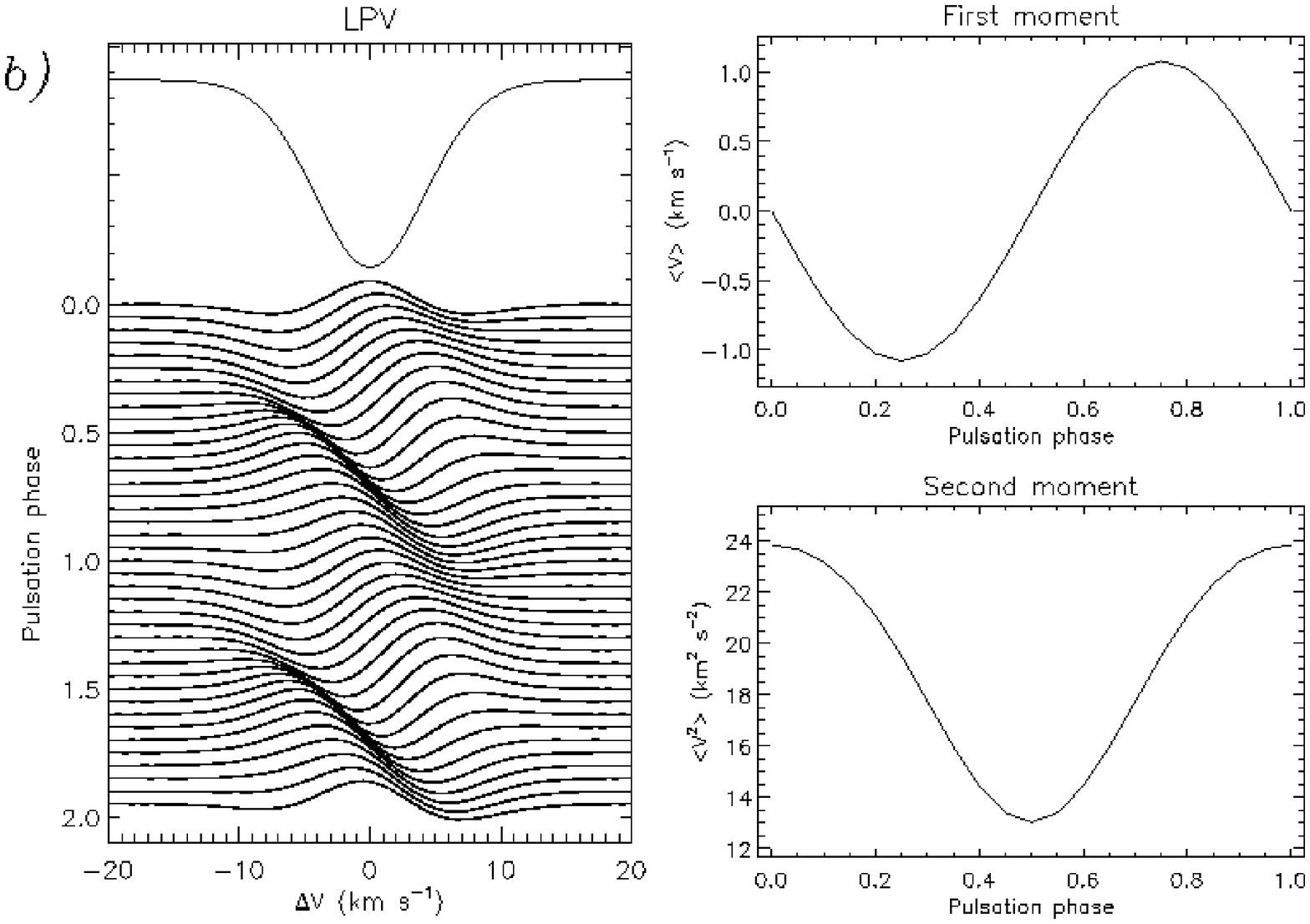}
\caption{Line profile variation of a non-rotating non-radial pulsator. {\bf a)} Spectrum
variability for the $\ell=1$, $m=0$, $V_p=2.0$~\kms, $K=0$ mode viewed from the pulsation pole. 
The local profile is represented by a constant Gaussian with FWHM
$W$\,=\,10~\kms. {\bf b)} Effect of adding harmonic variability of 
the line width with an amplitude $\delta W=1.5$~\kms\ and a phase shift of 0.25 with respect to the pulsational velocity variation. In both
cases the star is assumed to have $v_{\rm e}\sin i=0$~\kms. In each panel the left plot shows the average line profile on
top and time series  of the difference spectra (covering two pulsation cycles) below. The right
panels illustrates variation of the first (upper plot) and second (lower plot) line profile 
moments.} 
\label{fig13}
\end{figure*}

We found that none of the pulsation structures studied leads to a qualitative agreement with
observations. A typical failure of the theoretical model is illustrated in Fig.~\ref{fig13}a
for the $\ell=1$, $m=0$ mode viewed from the pulsation pole. The residual profile variation has a
symmetric S-shape and the line width changes have low-amplitude, double-wave character. In the
absence of stellar rotation, LPV for \textit{all} pulsation modes have these basic properties
-- the same conclusion follows from the theoretical analysis of the second moment behaviour
\citep*{APW92,K05}. This variability is not incompatible with observations of weak \ndt\ lines
in a few roAp stars, but clearly contradicts pulsation signatures found for the
majority of doubly ionized REE lines. 

When significant (\vs\,$\gg$\,$V_p$) rotational velocity is introduced in the model and
the pulsation pole of the dipolar mode is offset from the line of sight, one may encounter LPV
characterized by significant single-wave line width changes. However, the phase shift of RV and
line width oscillations is inconsistent with observations. Moreover, for this pulsational
geometry a small RV amplitude corresponds to a quite strong LPV, which has a fully symmetric
(both in time and relative to the line centre) H-shape and thus appears to be very different 
from the observed blue-to-red residual profile shift.

Our calculations show that all types of complex modes, obtained by combining different spherical
harmonics or by combining components with different phase shifts, also fail to achieve a satisfactory
agreement with observations. The key features of the roAp line profile behaviour -- single-wave
line width oscillations with a characteristic phase shift and asymmetric LPV pattern -- are
never found for the theoretical models explored. Thus, we conclude that the standard picture of
the velocity field in non-radial pulsators and trivial extensions of this picture are
inadequate for the REE lines which form in the upper atmospheres and show the most prominent
pulsation signatures in roAp stars. 

\begin{figure*}
\fifps{8.5cm}{0}{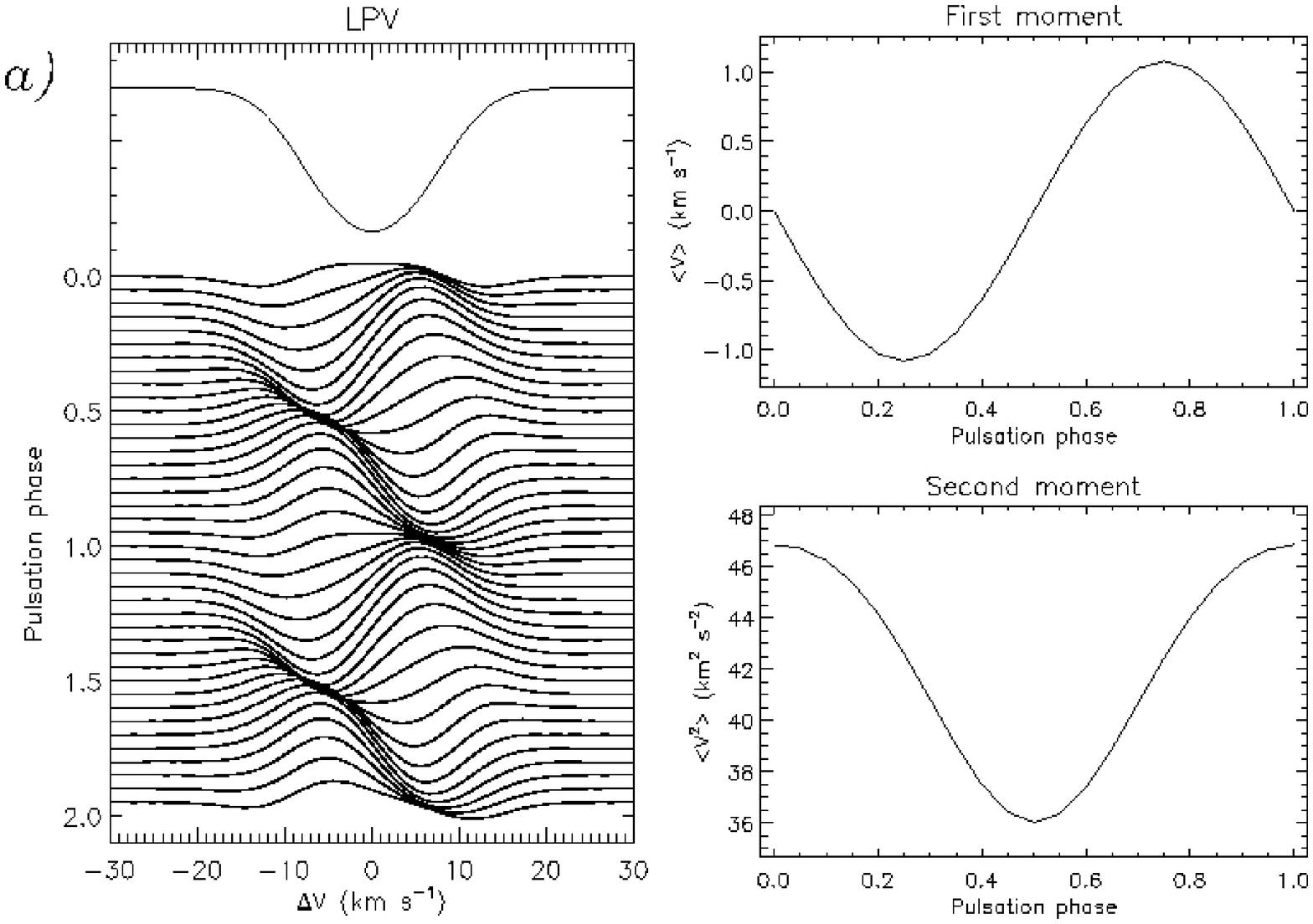}\hspace{0.2cm}
\fifps{8.5cm}{0}{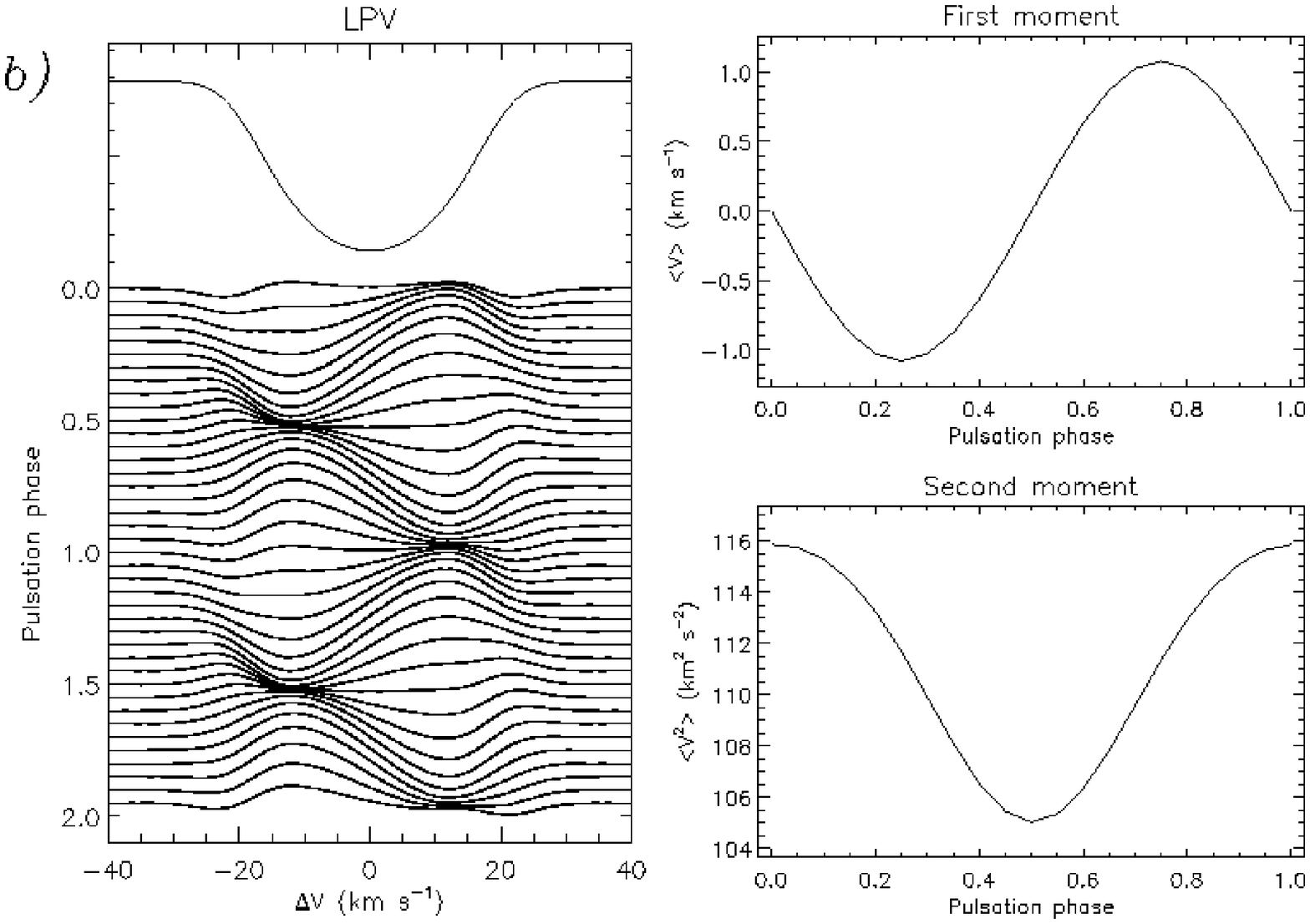}
\caption{The same as in Fig.~\ref{fig6}b for the oblique non-radial pulsator with 
{\bf a)} $v_{\rm e}\sin i=10$~\kms\ and {\bf b)} $v_{\rm e}\sin i=20$~\kms.}
\label{fig14}
\end{figure*}

Admittedly, the simplified line formation model adopted in our calculations does not allow one
to properly explore effects of the depth dependent pulsation amplitude and phase. However, we
expect that any superposition of several harmonic oscillations will give another harmonic
oscillation. Consequently, variation of the pulsation wave properties with height will not
introduce any qualitatively new behaviour and will not help to explain the puzzling spectroscopic observations of
roAp stars.

It appears that the key problem lies in the formulation of the standard pulsation framework:
the time dependent factor in Eq.~(\ref{vel}) postulates symmetry of the compression and
expansion parts of pulsation cycle. In reality the opposite, asymmetric behaviour is
observed in roAp stars (see Sect.~\ref{varia} and Fig.~\ref{fig10}).

\subsection{A new oblique pulsator model}

\begin{figure*}
\fifps{13.5cm}{0}{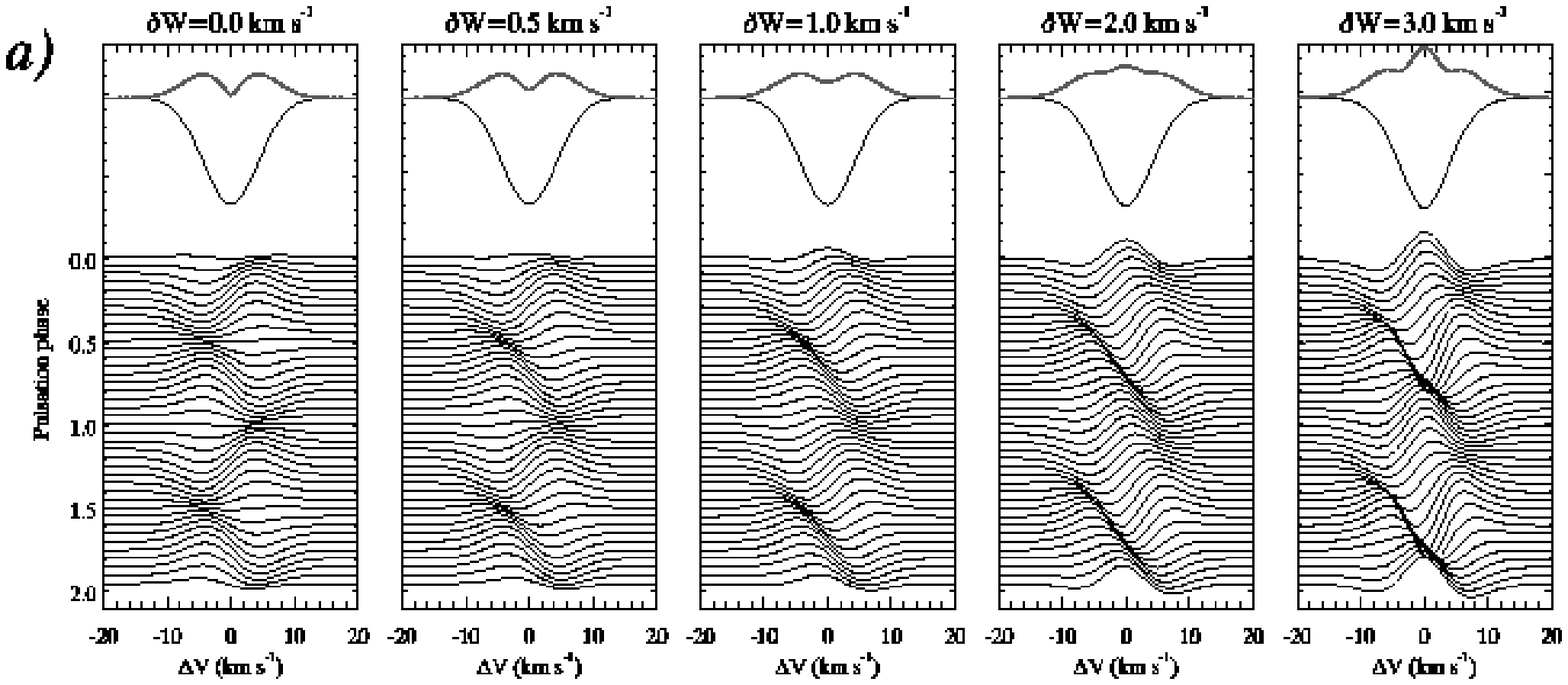}
\fifps{13.5cm}{0}{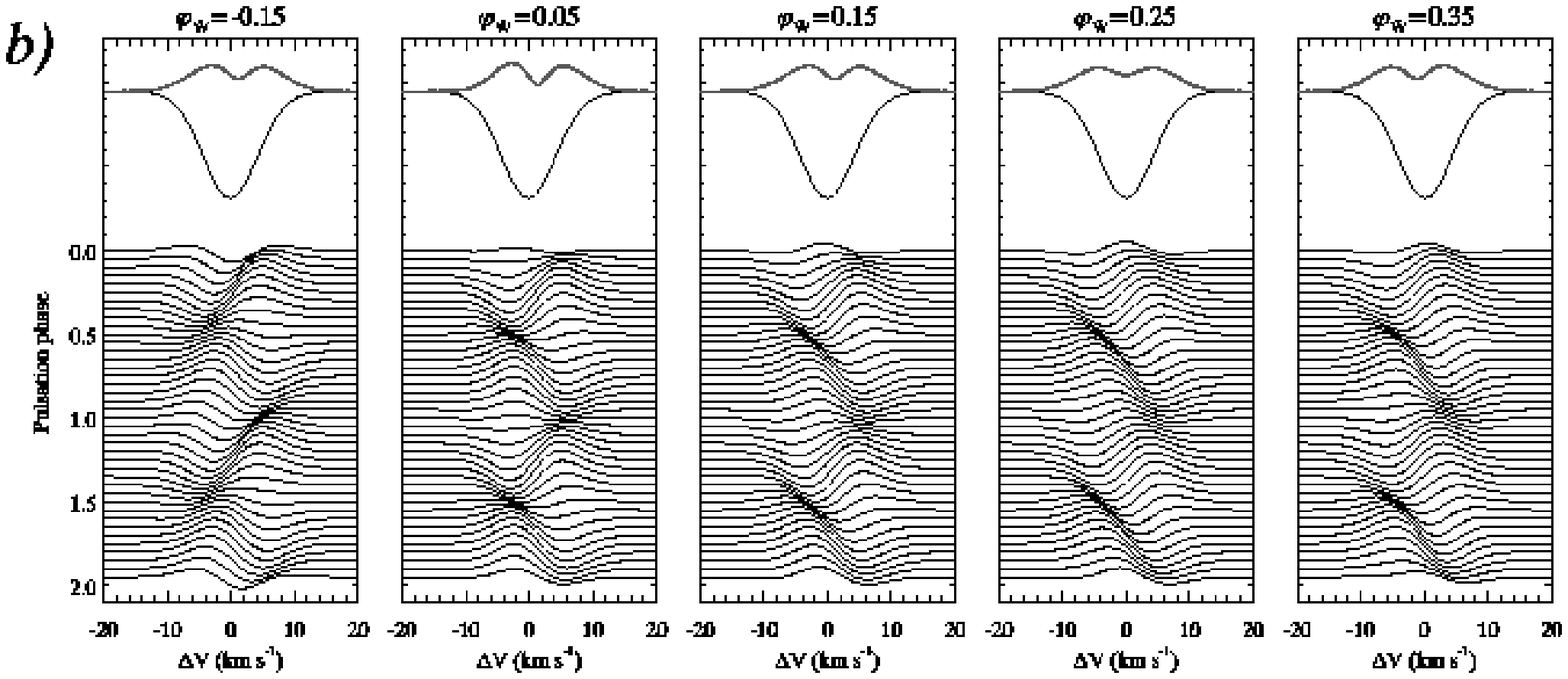}
\fifps{13.5cm}{0}{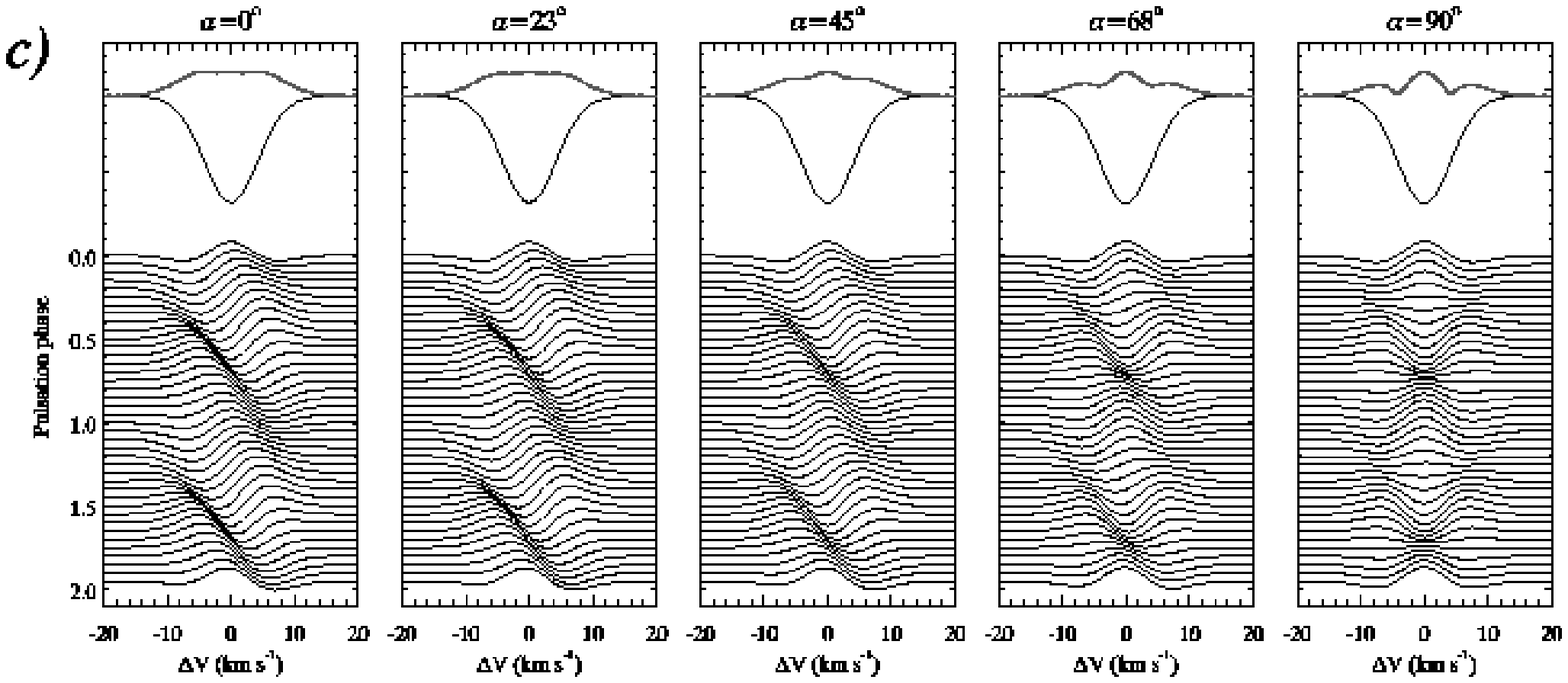}
\caption{Dependence of the line profile variation pattern on {\bf a)} amplitude $\delta W$ of the line 
width changes, {\bf b)} phase shift $\Delta\varphi$ between the line width and pulsation velocity 
variability, and {\bf c)} the angle $\alpha$ between the pulsation axis and the line of sight. The 
standard set of the oblique pulsator parameters is $\ell=1$, $m=0$, $V_p=2.0$~\kms, $K=0$, $\delta W=1.5$~\kms,
$\alpha=0\degr$. In each subpanel the average line profile and the standard deviation (arbitrary scale)
are presented on top. The time-resolved difference spectra, covering two pulsation cycles, are plotted 
below.}
\label{fig15}
\end{figure*}

We have discovered that the enigma of LPV in roAp stars can be resolved with a simple modification of the
classical oblique pulsator framework. Instead of looking for a combination of pulsation parameters
which gives the variation of the disk-integrated quantities illustrated in Fig.~\ref{fig10}, we suggest that
this empirical relation between radius, RV and second moment changes actually reflects a \textit{local}
periodic variation of velocity and line width. Thus, the key idea of our new oblique pulsator model is
that the pulsational mechanism responsible for the RV variability of REE lines gives rise to additional,
sinusoidal changes of line width. This variability occurs approximately in antiphase with the pulsational
radius changes and hence shows a quarter of period phase shift with respect to RV oscillations. Then the
LPV expected from this model is a combination of two effects:
\begin{enumerate}
\item the usual pulsational velocity changes described by spherical harmonic perturbations and
\item sinusoidal variability of line profile width, synchronized with changes of stellar radius.
\end{enumerate}

It is reasonable to assume that the surface distribution of the line width amplitude is similar to the 
vertical pulsational displacement and is described by
\beq
W(t,\theta,\phi) = W_0 + \delta W Y^m_\ell (\theta,\phi) {\rm e}^{({\rm i}\omega t-\varphi_W)}.
\eeq
The theoretical LPV can be calculated by incorporating this time and position-dependent line width in the
disk-integration formula (\ref{disk}). A striking effect of introducing variable line width in the line
profile synthesis is demonstrated by Fig.~\ref{fig13}b. The pulsation model used for this figure is
identical to the model of Fig.~\ref{fig13}a, except that now line width changes with an amplitude
$\delta W=1.5$~\kms\ around $W_0=10$~\kms\ and a phase shift $\varphi_W=0.25$. This leads to
a remarkable agreement between theoretical residual profile variation and observations of roAp stars.
Model spectra show smooth, blue-to-red moving pattern, which is very similar to the LPV discussed in
Sect.~\ref{resid}. The additional variation of line width does not distort the sinusoidal behaviour of the RV.
Considering the second moment oscillations, we find that a large-amplitude single-wave variability
superimposed on top of a small-amplitude double-wave curve results in nearly sinusoidal variation of 
\vb. The phase shift assumed for the time dependence of the local velocity and line width 
propagates to the disk-integrated observables, allowing us to reproduce the observed phase lag of \vb\
with respect to \va.

An interesting modification of the LPV pattern calculated with our model is associated with increasing
\vs. The LPV corresponding to \vs\,=\,0~\kms\ is presented in Fig.~\ref{fig13}b. Fig.~\ref{fig14}a and
b show LPV for the same oblique pulsator geometry and amplitude, but with \vs\ increased to 10 and
20~\kms, respectively. The blue-to-red shift is still discernible for the \vs\,=\,10~\kms\ model,
although it is not as smooth as in nonrotating star. In the case of rapid rotation
(Fig.~\ref{fig14}b), the signature of line width variability disappears and the LPV shows an almost symmetric
pattern, as one would expect for the standard oblique pulsator. Thus, if the $\delta W$ amplitude
adopted for these simulations is realistic, we predict that a signature of the pulsational variation
of line width will be visible only in slowly rotating stars (\vs\,$\la$\,10~\kms), but not in
relatively rapid rotators (\vs\,$\ga$\,20~\kms). In fact, this prediction is verified by observations.
We find asymmetric LPV in all slowly rotating roAp stars, including HD\,9289 and HD\,128898
(\vs\,$\sim$\,10~\kms), whereas the study of high-amplitude, rapidly rotating (\vs\,$=$\,33~\kms) 
roAp pulsator HD\,83368 reveals a symmetric LPV pattern and no evidence of blue-to-red shifting
features in residual line profiles \citep{K06}.

We have performed additional calculations with the aim of examining the sensitivity of the theoretical LPV to
the parameters of the new oblique pulsator model. Fig.~\ref{fig15} illustrates the modifications of
profile variations due to changes of the amplitude and phase of the line width oscillations and due
to changes of the angle $\alpha$ between the line of sight and pulsation axis. Gradual increase of
the line width amplitude transforms symmetric, S-shaped LPV into asymmetric, blue-to-red pattern
(see Fig.~\ref{fig15}a). Thus, the diversity and height dependence of the LPV observed in roAp stars can
be plausibly attributed to different $V_p$--$\Delta W$ combinations and to a growth of the line
width amplitude with height.  Qualitative comparison of the simulated profile time series and
observations suggests that for most stars $\delta W$ lies in the range from 0 to 1.5--2~\kms.
Fig.~\ref{fig15}a also demonstrates that the shape of the standard deviation profile changes from a
double peak to single peak when $\delta W$ is increased. This agrees with our observations of the
correlation between LPV pattern and the standard deviation shape. The fact that we never observe a
dominant central peak in the standard deviation signatures also points to $\delta W$ amplitudes
below 1.5--2~\kms.

Changing the $\varphi_W$ parameter from 0 to $\approx0.15$ also strongly affects the LPV character.
Fig.~\ref{fig15}b clearly shows that it is a non-zero phase shift between the velocity and line width
oscillation that introduces an asymmetry in LPV. The blue-to-red shifting features are obtained for
positive $\varphi_W$ exceeding $\approx0.1$. On the other hand, negative $\varphi_W$ results in the
opposite, red-to-blue, pattern, possibly detected in \ndt\ lines of HD\,9289. The morphology of the LPV is
relatively unaffected by variation of the phase shifts between 0.15 and 0.35. Interestingly, a $\varphi_W$
different from 0 and 0.25 produces an asymmetric standard deviation, which is sometimes shifted with
respect to the line centre. This provides the first explanation of the origin of distorted and shifted
standard deviation profiles frequently observed in slowly rotating roAp stars.

Rotational modulation of the LPV pattern presented in Fig.~\ref{fig15}c shows a small change of
profile variation as the line of sight shifts from the pulsation pole to the pulsation equator of the
oblique dipolar mode. A dramatic change of the LPV is associated with a narrow range of rotation phases
around $\alpha=90\degr$, when the mode is viewed equator-on. For this pulsation geometry the RV
variation is negligible and LPV is caused entirely by the line width oscillations.

The line profile variation calculations presented in this section convincingly demonstrate the success
of our new oblique pulsator model. Simultaneous pulsational variation of velocity and line width
naturally explains a wide range of the LPV behaviour observed in roAp stars. We showed that, by
adjusting a few parameters (amplitude and phase of the line width variation) of our model, it becomes
possible to reproduce the diverse variations found for different groups of REE lines.

\begin{figure}
\figps{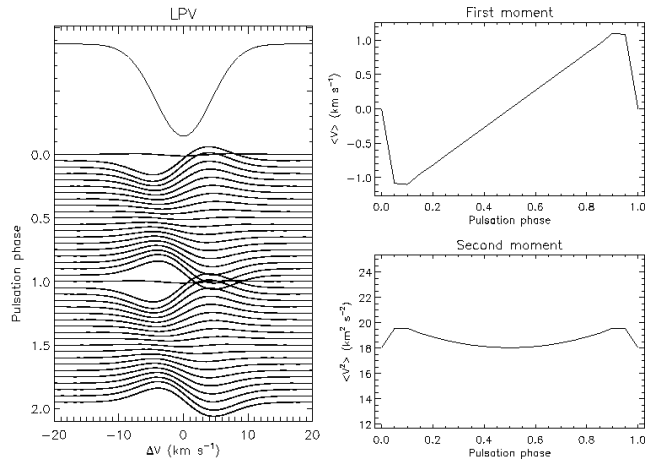}
\caption{Line profile variation expected for the oblique non-radial pulsator within the framework
of the shock wave model proposed by \citet{SKK04}. The velocity field geometry
is that of the pole-on $\ell=1$, $m=0$, $V_p=1.6$~\kms\ pulsation mode. The temporal variation
is given by the smoothed saw-tooth function $\sum_{n=1}^{10}n^{-1}\exp{({\rm i} n\omega t)}$. Other parameters and
the layout of the figure are the same as in Fig.~\ref{fig13}a.}
\label{fig16}
\end{figure}

\subsection{A shock wave signatures in the LPV of roAp stars?}

Recently, another interpretation of pulsational profile variations in roAp stars was proposed by
\citet{SKK04}. These authors suggested that the smooth, blue-to-red shift observed by \citet{KR01a} in
the LPV of the \nd\ 6145~\AA\ line in \equ\ represents the signature of shock waves propagating in the
high atmospheric layers. \citeauthor{SKK04} deduce the velocity of the pulsation wave, not from the amplitude
of the RV oscillation ($\approx$\,500~\ms\ for this \nd\ line), but from the full span of the residual profile
variation (see Fig.~\ref{fig1}). This gives a maximum speed of $\sim$\,18~\kms, which is substantially
higher than the sound speed. Based on these arguments, \citeauthor{SKK04} claim that LPV is affected
by shocks and the harmonic time dependence in Eq.~(\ref{vel}) should be replaced by a saw-tooth law
typical for large-amplitude non-linear stellar oscillations.

We believe that the basic assumption of \citet{SKK04} is incorrect because it is based on an erroneous
interpretation of the pulsational profile distortions in terms of the physical velocity of the 
pulsation wave. The velocity span of the residual profile variation is determined by a convolution of the intrinsic
stellar line profile with the pulsation velocity field. If the former is large compared to the pulsation
velocity amplitude, as is the case for REE lines in roAp stars, the velocity span of LPV is not
representative of the true pulsation wave speed. This situation is analogous to the behaviour of
rapidly rotating non-radial pulsators, where profile variation with $V_p\ll$\,\vs\ is observed in the
whole velocity range from $-$\vs\ to $+$\vs\ \citep{STA97}. Thus, we assert that observations of \equ\
and other roAp stars give no evidence of supersonic pulsation velocity speeds and therefore the model
of \citet{SKK04} cannot be correct.

However, it is still instructive to investigate the consequences of a non-sinusoidal velocity variation for
oblique pulsators. In doing this, we have followed \citet{SKK04} and replaced the 
$\exp{({\rm i} \omega t)}$ term of Eq.~(\ref{vel}) by $\sum_{n=1}^{10}n^{-1}\exp{({\rm i} n\omega t)}$.
Additional moderate smoothing along the time axis is required to get rid of wiggles introduced by a
finite number of terms in the latter expression. The line profile and moment variability resulting from a
saw-tooth time dependence of the pulsation velocity is presented in Fig.~\ref{fig16}. Evidently, the residual
LPV shows a blue-to-red shift, which is superficially similar to the observed profile behaviour.
However, the time series of residuals also shows a prominent gap due to the very fast evolution of the profile
shape at phase 0. This is essentially the signature of the shock wave postulated by \citet{SKK04}. No
such rapid changes of line profiles have been reported for any roAp star. Also, in contradiction with
observations, the variation of the line profile moments predicted by the shock wave model contains a large
harmonic contribution. The observed phase relation of RV and line width is not reproduced either: the
maximum line width coincides with the velocity extremum, whereas observations point to a 0.25 phase offset
between the two curves.

In summary, our theoretical calculations suggest that, even putting aside the dubious foundation of
the shock wave model, the line profile and moment variability predicted by this theory disagrees with
observations. In our opinion, this proves that the model formulated by \citet{SKK04} does not
provide a viable solution to the puzzle of LPV in roAp stars. 

\section{Discussion}
\label{disc}

Our survey of pulsational line profile variability has, for the first time, characterized in
detail the rapid fluctuations of REE lines in a representative sample of slowly rotating roAp stars. For a
subset of lines in each star studied, we demonstrated the presence of asymmetric LPV of the type first
reported by \citet{KR01a} for \equ. We point out the close relation of this variability to the
high-amplitude oscillation of the second moment. Both the asymmetric, blue-to-red shifting features in
the residual profile time series, and the single-wave line width modulation with pulsation phase,
increase rapidly with height and reach a maximum in the uppermost atmospheric layers, probed by strong
doubly ionized REE lines. We show that this type of LPV is incompatible with the standard oblique
pulsator model, which attributes pulsational spectroscopic variability to the velocity perturbations
described by the $\ell=1$, $m=0$ spherical harmonic. Extensive line profile calculations reveal that
none of the recent modifications and extensions of the oblique pulsator framework
\citep{BD02,SG04,C06}, nor the exotic shock wave model of \citet{SKK04}, are capable of explaining the
LPV observed in sharp-lined roAp stars. Neveretheless, we show that theoretical line profile variation
can be brought into remarkable agreement with experimental data if a pulsational modulation of the width
of the local intrinsic spectral line profile is incorporated into the model. This new oblique pulsator
framework is based on dipolar modes, and thus does not contradict observational evidence of the
presence of this type of non-radial oscillations, accumulated by numerous previous spectroscopic and
photometric studies of roAp stars. At the same time, a combination of the velocity and line width
variability solves the puzzle of asymmetric LPV, explains why this line profile behaviour is clearly
detected only in sharp-lined roAp stars, and provides a satisfactory interpretation of the line profile
moments and the morphology of the standard deviation spectrum. The empirical success of our line width
modulation hypothesis notwithstanding, the important question of the physical mechanism responsible for
the pulsational alternation of the width of spectral lines has not yet been addressed. Here we attempt to
shed light on this problem.

\begin{figure}
\figps{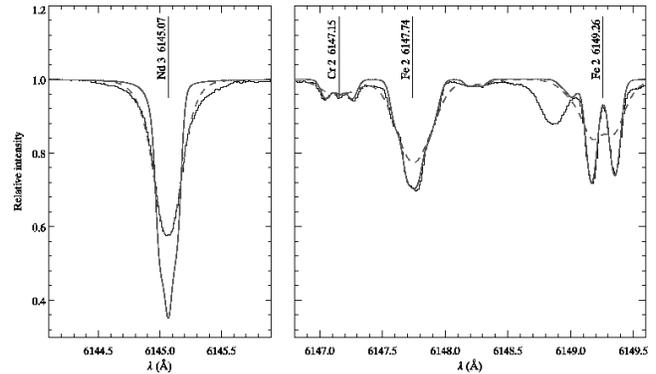}
\caption{Comparison of very high resolution observation of \equ\ (thin curve) with magnetic
spectrum synthesis calculations. Zeeman resolved profiles of the iron-peak lines (right panel)
are well reproduced with a macroturbulent broadening of 2~\kms\ (thick solid curve). In
contrast, a satisfactory fit to the broad wings of the \nd\ 6145~\AA\ line (left panel) is 
only possible for $V_{\rm macro}\ge 10$~\kms\ (dashed curve).}
\label{fig17}
\end{figure}

High-resolution, high-$S/N$ observations of cool Ap stars reveal a number of line profile and line
strengths anomalies, usually attributed to the combined effects of chemical stratification, a strong
magnetic field, and non-solar chemical composition \citep{RPK02}. However, the significant broadening of
strong doubly ionized REE lines in slowly rotating cool Ap stars has received no satisfactory explanation
so far. This problem is illustrated in Fig.~\ref{fig17} for the \nd\ 6145~\AA\ line in \equ. We
compare the highest resolution observation of this star available in the ESO Archive (coadded
$\lambda/\Delta\lambda=220,000$ spectra obtained by \citet{SHM05}) with the magnetic spectrum
synthesis calculations based on the stellar parameters, chemical abundances and magnetic field model
of \citet{RPK02}. The additional broadening required to fit the Zeeman resolved components of the \ion{Fe}{ii}
and \ion{Cr}{ii} lines is approximately 2~\kms. This is much smaller than $V_{\rm macro}\ge 10$~\kms\
required to roughly reproduce the profile of the \nd\ line. Such a high Gaussian smearing washes out the Zeeman
structure in the Fe-peak lines and, therefore, contradicts observations of spectral lines formed in the lower
atmosphere of \equ. The factor of five discrepancy in the best-fitting macroturbulence derived for
iron-peak elements and \nd\ cannot be ascribed to isotope splitting or hyperfine structure effects
(those are small for \nd, see \citet{DWL02}) and seems to reflect a genuine increase of the isotropic
velocity component towards the high-lying REE cloud.

Unusually strong broadening of pulsating REE lines was first discussed for \equ\ by \citet{KR01a}. In
that paper we advocated a high-$\ell$, non-axisymmetric mode in \equ\ and attributed the large broadening
of \pr\ and \nd\ lines to the unresolved, large-amplitude pulsational velocity field. Our new
investigation of this and other roAp stars suggests a different picture: pulsation modes are
axisymmetric with $\ell=1$,  and line width modulation is invoked to explain the observed LPV. This model implies
local pulsational perturbations with amplitudes not exceeding 1--2~\kms\ for the REE line-forming
layer, which leaves no room for $\sim$\,10~\kms\ pulsational broadening in the time-averaged spectrum.
Comparison of the pulsational amplitude and line broadening for different time-resolved spectroscopic
data sets of \equ\ and for different roAp stars also argues against a direct relation between  the average
line width and non-radial pulsation. We also find no differences between the REE line broadening for time-averaged
spectra obtained from time series of \equ\ characterized by an overall RV amplitude that is different by a factor
of five. Many roAp stars have relatively low ($<$\,100~\ms) pulsation amplitudes (HD\,137949,
HD\,166473, HD\,176232) but, nevertheless, show very broad REE lines. Finally, apart from the small
line width modulation discussed in this paper, no systematic reduction of the REE line widths is
observed in time-resolved spectra.

Thus, effects other than pulsational variability must be responsible for the inexplicable broadening
of REE lines. Turbulence caused by convective instability is an obvious candidate for explanations of
roAp-star observations. The location of cool Ap stars in the H-R diagramm coincides with Am stars, for
which large broadening of strong lines is observed and interpreted as a signature of convective
motions \citep{L98}. On the other hand, the intense magnetic fields of Ap stars are likely to modify and,
possibly, suppress H and He convection zones. Ubiquitous signatures of chemical stratification also
indicate a lack of convective mixing in the lower atmosphere of cool Ap stars, though some mixing is
required to reduce the discrepancy between theoretical diffusion models and observations \citep{LM04}. The
message of these modern theoretical and observational results is that magnetic field-convection
interaction in cool Ap stars is far more complex than previously throught and that a common assumption
of the full suppression of convection is likely to be an oversimplification.

With this background, turbulence related to convective or pulsationally-driven instability is the most
plausible explanation of the anomalous broadening of REE lines. We speculate that some cool Ap stars,
including the roAp pulsators studied in our paper, possess a turbulence zone in the upper atmospheric
layers, roughly at the height of the REE-rich cloud. Large vertical gradients of chemical composition
in the intermediate atmosphere ($-3.5$\,$\le$\,$\log\tau_{5000}$\,$\le$\,$-0.5$) exclude significant mixing.
Therefore, turbulence should appear only in layers above $\log\tau_{5000}$\,$\approx$\,$-3.5$ to $-4.0$. We
find that strong REE lines often display excessive width of the outer wings, which is difficult to
reproduce assuming a height-independent broadening parameter. Possibly, line cores sample layers with
a weaker turbulence -- a situation consistent with a decrease of convection efficiency with height.

At this stage we can offer no complete physical theory explaining the appearance of the turbulence zone in
the upper atmospheres of cool Ap stars. One reason for the convective instability to develop may be a
non-standard temperature structure of Ap-star atmospheres. Empirical evidence for this kind of anomaly
was found by \citet*{KBB02}. By fitting the peculiar shape of the core-wing transition of the hydrogen
Balmer lines, they derived semi-empirical models, characterized by a nearly flat temperature profile
above $\log\tau_{5000}$\,$\approx$\,$-1.0$ followed by a steep drop of temperature at
$\log\tau_{5000}$\,$\approx$\,$-4.0$. This large temperature gradient may lead to convective
instability\footnote{We thank Dr. H. Saio for drawing our attention to this fact.} very close to the
inferred location of the turbulence zone needed to explain broadening of REE lines.

What could be the response of the hypothetical turbulent layer to the non-radial $p$-mode oscillations?
We know that the amplitude of convective motions grows as the mean atmospheric pressure increases
\citep{H02}. Pulsational modulation of convection efficiency was invoked to explain observations  for
several classes of pulsating stars \citep{FGB96,GDF98}. One usually infers that convective motions
become stronger during the compression phase of the oscillation cycle. Conversly, convection is less
intense during the expansion of the pulsating star. This is precisely the relationship between the line width and
radius variation that we have deduced for roAp stars. Consequently, it is reasonable to suggest that
pulsations modulate the efficiency of turbulence in the upper atmospheres of roAp stars, and this leads to
the observed line width and line profile variation.

We hope that our interpretation of the pulsational LPV in roAp stars and the possible discovery of a new
turbulence zone will stimulate new detailed magnetohydrodynamic modelling of the interaction between
convection, pulsation and magnetic field in cool Ap stars. Three-dimensional radiative hydrodynamic
simulations of A-star convection \citep[e.g.,][]{KFP06} should be extended to include large-scale
fossil magnetic fields. Variable lower boundary condition can be used to simulate the effects of
high-overtone $p$-modes. On the other hand, a NLTE semi-empirical analysis of hydrogen lines, coupled
with model atmosphere calculations taking into account chemical stratification and the magnetic field
\citep{STR04,KKS05}, is necessary to verify and extended the results of \citet*{KBB02}.

\section*{Acknowledgments}
This paper is based on observations obtained at the European Southern Observatory (Paranal, Chile)
and at the Canada-France-Hawaii Telescope.
Resources provided by the electronic databases (VALD, SIMBAD, NASA's ADS)
are acknowledged. Our research is supported by the grants from the Swedish 
Kungliga Fysiografiska S\"allskapet, Swedish Royal Academy of Sciences (project 11630102), 
Russian Foundation for Basic Research (project 04-02-16788a),
the Austrian Science Fund (project P17580), and the Natural Sciences and Engineering Research Council of Canada.

\bsp

\end{document}